\documentclass[journal,10pt]{IEEEtran}
\usepackage[T1]{fontenc}
\usepackage{amsmath,bm,amsfonts,amssymb,graphicx,color,multirow,setspace,xfrac,comment,xcolor}
\usepackage{booktabs}
\usepackage{epstopdf}
\usepackage{mathtools}
\usepackage{lipsum}
\usepackage{cite,citesort}
\usepackage{float}
\usepackage{balance}
\usepackage{gensymb}
\newcommand{\tabitem}{~~\llap{\textbullet}~~}

\begin{document}
\title{6G Wireless Systems: Vision, Requirements, Challenges, Insights, and Opportunities\vspace{5pt}}
\author{Harsh~Tataria,~\IEEEmembership{Member,~IEEE,}
       Mansoor~Shafi,~\IEEEmembership{Life~Fellow,~IEEE,}
       Andreas~F.~Molisch,~\IEEEmembership{Fellow,~IEEE,} 
       Mischa~Dohler,~\IEEEmembership{Fellow,~IEEE,}
       Henrik~Sjöland,~\IEEEmembership{Senior Member,~IEEE,} 
       and Fredrik~Tufvesson,~\IEEEmembership{Fellow,~IEEE}
\vspace{-19pt}
\thanks{H.~Tataria, H.~Sjöland, and F.~Tufvesson are with the Department of Electrical and Information Technology, Lund University, Lund, Sweden (e-mail: \{harsh.tataria, henrik.sjoland, fredrik.tufvesson\}@eit.lth.se). }
\thanks{M.~Shafi is with Spark New Zealand, Wellington, New Zealand (e-mail: mansoor.shafi@spark.co.nz).}
\thanks{A.~F.~Molisch is with the Ming Hsieh Department of Electrical and Computer Engineering, University of Southern California, Los Angeles, CA, USA (e-mail: molisch@usc.edu).}
\thanks{M.~Dohler is with the Center for Telecoms Research, King's College London, London, UK (e-mail: mischa.dohler@kcl.ac.uk).}
\thanks{The work of H.~Tataria, H.~Sjöland, and F.~Tufvesson was supported by Ericsson AB, Sweden, and ELLIIT: The Linköping-Lund Excellence Center on IT and Mobile Communication. The work of A.~F.~Molisch was supported by the National Science Foundation (NSF), the National Institute of Standards and Technology (NIST), and Samsung Research America.}
\vspace{-9pt}}
\maketitle

{\color{black}
\begin{abstract}
Mobile communications have been undergoing a generational change every ten years or so. However, the time difference between the so-called ``G's" is also decreasing. While fifth-generation (5G) systems are becoming a commercial reality, there is already significant interest in systems beyond 5G, which we refer to as the sixth-generation (6G) of wireless systems. In contrast to the already published papers on the topic, we take a \emph{top-down} approach to 6G. More precisely, we present a holistic discussion of 6G systems beginning with lifestyle and societal changes driving the need for next generation networks. This is followed by a discussion into the technical requirements needed to enable 6G applications, based on which we dissect key challenges, as well as possibilities for practically realizable system solutions across \emph{all} layers of the Open Systems Interconnection stack (i.e., from applications to the physical layer). Since many of the 6G applications will need access to an order-of-magnitude more spectrum, utilization of frequencies between 100 GHz and 1 THz becomes of paramount importance. As such, the 6G eco-system will feature a diverse range of frequency bands, ranging from below 6 GHz up to 1 THz. We comprehensively characterize the limitations that must be overcome to realize working systems in these bands; and provide a unique perspective on the physical, as well as higher layer challenges relating to the design of next generation core networks, new modulation and coding methods, novel multiple access techniques, antenna arrays, wave propagation, radio-frequency transceiver design, as well as real-time signal processing. We rigorously discuss the fundamental changes required in the core networks of the future, such as the redesign or significant reduction of the transport architecture that serves as a major source of latency for time-sensitive applications. This is in sharp contrast to the present hierarchical network architectures, which are not suitable to realize many of the anticipated 6G services. While evaluating the strengths and weaknesses of key candidate 6G technologies, we \emph{differentiate} what may be practically achievable over the next decade, relative to what is possible in theory. Keeping this in mind, we present concrete research challenges for each of the discussed system aspects, providing inspiration for what follows. 
\vspace{-15pt}
\end{abstract}


\vspace{-15pt}
\section{Introduction}
\label{Introduction}
\vspace{-2pt}
Enabled by enhanced mobile broadband (eMBB), new applications in massive machine type communications (mMTC) and ultra reliable low-latency communications (uRLLC) has driven the development towards International Mobile Telecommunications 2020 (IMT-2020) - often colloquially called the fifth-generation (5G) of wireless systems \cite{shafi20175g,7403840}. As the next decade unfolds, extremely rich multimedia applications in the form of high-fidelity holograms and immersive reality, tactile/haptic-based communications, as well as the support of mission critical applications for connecting all things are being discussed \cite{OULUWP1,7403840}. To support such applications, even larger system bandwidths than those seen in 5G are required along with new physical layer (PHY) techniques, as well as higher layer capabilities which are not present today. Significant efforts are underway to characterize and understand wireless systems beyond 5G, which we refer to as the sixth-generation (6G) of systems \cite{OULUWP1,ZHANG6G1,khaled,Li,FG2030}. Research on 6G wireless systems is now the center of attention for a large number of journal and conference publications, keynote talks and panel discussions at flagship conferences/workshops, as well as in the working groups of standardization bodies, such as the International Telecommunications Union-T (ITU-T) \cite{OULUWP1,FG2030,SAMSUNGWP1}. For the vast majority of these studies, the scope of the work ranges from characterizing potential 6G use cases, identifying their requirements, and analyzing possible solutions - in particular for PHY of the Open Systems Interconnection (OSI) stack. 

Nevertheless, in order to understand what future systems will be capable of, we first provide details on evolving requirements of daily life approaching the next decade, which will naturally drive the requirements for 6G. To this end, we summarize the key drivers behind 6G systems, discuss the literature summarizing the 6G vision as well as performance metrics, and present the contributions of this paper. Followed by this, we present the organization of the remaining sections of the paper. 

\vspace{-10pt}
\subsection{Drivers for 6G Systems: Lifestyle and Societal Changes}
\vspace{1pt}
\label{KeyDriversfor6GSystemsLifestyleandSocietalChanges}
According to the ITU-T in \cite{FG2030}, the \emph{three} most important driving characteristics linked to the next decade of lifestyle and societal changes, impacting the design and outlook of 6G networks are: \emph{1) High Fidelity Holographic Society}, \emph{2) Connectivity for All Things}, and \emph{3) Time Sensitive/Time Engineered Applications}. In what follows, we present our view of each disruptive change and connect its implications to wireless networks of the future.

\subsubsection{\underline{High Fidelity Holographic Society}}
\label{HighFidelityHolographicSociety}
Video is increasingly becoming the mode of choice for communications today, and is evolving to augmented reality. As such, video resolution capability is increasing at a rapid rate. For instance, user equipment (UE) devices supporting 4K video require a data rate of 15.4 Mbps (per-UE) \cite{shafi20175g}. 
Additionally, a UE's viewing time is also increasing, to the point where it is now the norm for end-users to watch complete television programs, live sports events, or on-demand streaming. As we enter the next decade, demand for such content is anticipated to grow at extreme rates \cite{OULUWP1,SAMSUNGWP1}. The ongoing COVID-19 pandemic is showing that video communication has enabled people, businesses, governments, medical professionals and their patients to remain in virtual contact, avoiding the need for travel  while remaining socially, professionally, and commercially active. While educational institutions remain closed, online education is possible via video communication. At the time of writing this paper, premier conferences and workshops around the world are being held virtually using live video interfaces. We expect that many such developments will remain active, even in the post COVID-19 era. 

Holograms and multi-sense communications are the next frontier in this virtual mode of communication. In 2017, the renowned physicist Stephen Hawking gave a lecture to an audience in Hong Kong via a hologram showcasing the growing potential of such a technology. Holograms are not just a technological gimmick or limited to entertainment; rather a logical evolution of video communication providing a much richer user experience. Proof-of-concept trials of hologrammatic telepresence are already underway \cite{gotsch}. When it is deployed, holographic presence will enable remote users as a rendered local presence. For instance, technicians performing remote troubleshooting and repairs, doctors performing remote surgeries, as well as improved remote education in classrooms could benefit from hologram renderings. The data transmission rates for holograms are very substantial (at least for today). Besides the standard video properties such as colour, depth, resolution, and frame rate, holographic images will need transmission from \emph{multiple} view points to account for variation in tilts, angles and observer positions relative to the hologram. As an example, if a human body is mapped in tiles, say of dimensions 4"$\times$4", then a 6'$\times$20" person may need a transmission rate of 4.32 Tbps \cite{Li}. \emph{This is substantially more than what 5G systems are capable of providing.} In addition, to consistently provide such high data rates, additional \emph{synchronization} is required to coordinate transmissions from the multiple view points ensuring seamless content delivery and user experience. Some applications may need to combine holograms with data from other sources. This would enable data to be fed back to a rendered entity from a remote point. Combinations of tactile networks and holograms, especially if we are able to provide \emph{touch} to the latter, could open further applications. 

While audio, video and holograms involve the senses of sight and hearing, communication involving \emph{all} of the \emph{five senses} is also being considered. Smell and taste are considered as lower senses, and are involved with feelings, as well as emotions; thus digital experiences can be enriched via smells and tastes. In general, we believe that a variety of sensory experiences may get integrated with holograms. To this end, using holograms as the medium of communication, emotion-sensing wearable devices capable of monitoring our mental health, facilitating social interactions and improving our experience as users will become the building blocks of networks of the future \cite{ShuChiu}.

\subsubsection{\underline{Connectivity for All Things}}
\label{ConnectivityforAllThings}
Using 5G as a platform, an order-of-magnitude or even higher number of planned inter-connectivity, and its widespread use will be another defining characteristic of the future society. This will include infrastructure that is essential for the smooth functioning of society that we have become used to today, such as water supplies, agriculture, uninterrupted power, transport and logistics networks, etc. This brings the necessity to operate \emph{multiple network types}, going well beyond the standard terrestrial networks of today. There are significant attempts to develop uninterrupted global broadband access via integration between the terrestrial networks and many planned satellite networks - especially for low earth orbit (LEO) satellites. Communication from moving platforms, such as Unmanned Ariel Vehicle (UAV)-based systems is also required as many new applications are emerging. In addition to this, there is also a desire to explore life on other planets. Successful operation of such critical infrastructure brings the need for \emph{security} beyond what is possible today. In addition to this, the increased reliability of the sensors monitoring the infrastructure is also essential to successfully migrate towards a truly connected society.

\subsubsection{\underline{Time Sensitive/Time Engineered Applications}}
\label{TimeSensitiveandTimeEngineeredApplications}
Humans and machines are both sensitive to delays in the delivery of information (albeit to varying degrees). \emph{Timeliness} of information delivery will be critical for the vastly interconnected society of the future. New applications that intelligently interact with the network will demand guaranteed capacity and timeliness of arrivals. As we incorporate gadgets in our life, quick responses and real-time experiences are going to be increasingly relevant. In a network of a massive number of connected sensors which are the end points of communication, timeliness becomes critical and late arrival of information may even be catastrophic.\footnote{To take the example of an autonomous car, the large numbers of other vehicles, pedestrians, traffic signals, street signs and other identifiable objects may become the communication end points.} Time sensitivity also has a deep impact on other modes of communications in the future, such as those relying on \emph{tactile and haptic control}. Conventional internet networks are capable of providing audio and video facilities, which can be classified as non-haptic control of communication. However, the tactile internet \cite{Fettweis,Aijaz} will also provide a platform for touch and actuation in real-time. Due to the fundamental system design and architectural limitations, current 5G systems are not able to completely virtualize any skill performed in another part of the world, and transport it to a place of choice, under the 1 ms latency limit of human reaction. This will be addressed in 6G systems with leaner network architectures and more advanced processing (see e.g., \cite{Aijaz} and references therein). 

\vspace{1pt}
With the above changes driving the need for 6G, we review the progress in literature on 6G systems. We note that besides the studies referred to in the subsection below, there are many papers dealing with specific technologies at the PHY, media access control (MAC), and transport layers of the OSI stack. These papers will be reviewed (partly) in the related sections of this paper. \emph{Overall, we stress that since 6G encompasses a large part of ongoing communications research, any literature review is necessarily incomplete and can only provide important examples.}

\vspace{-10pt}
\subsection{Literature Review: 6G Vision and Performance Aspects}
\label{LiteratureSurvey6GVisionandPerformanceAspects}
By now, a considerable number of papers have explored possible applications and solutions for 6G systems. For instance, the authors of \cite{giordani} take a look at potential 6G use cases, and provide a system-level perspective on 6G requirements, as well as presenting potential technologies that will be needed to meet the listed requirements. The studies in \cite{SAMSUNGWP1,OULUWP1,gui20206g} give a flavour of the possible key performance indicators (KPIs) of 6G systems, and provides a summary of enabling technologies needed to realize the KPIs, such as holographic radio (different from standard holograms), terahertz (THz) communications, intelligent reflecting surfaces (IRS) and orbital angular momentum (OAM). On a similar theme, the authors of \cite{bariah2020prospective,chen2020vision,tariq,yuan2019potential,chen} present the applications and enabling technologies for 6G research and development.

A number of studies focusing on more specific technologies have also been published. For instance, the study in \cite {corre2019sub} proposes to explore new waveforms for 90-200 GHz frequency bands that offer optimal performance under PHY layer impairments. The authors of \cite{haselmayr2019integration} present a vision of providing an internet of bio-nano things using molecular communication. The study in \cite{lopez2019massive} gives an overview of architectures, challenges and techniques for efficient wireless powering of internet-of-things (IOT) networks in 6G. Moreover, the authors in \cite{piran2019learning} consider the requirements, use cases and challenges to realize 6G systems with a particular emphasis on \emph{artificial intelligence (AI)}-based techniques for network management. The role of \emph{collaborative} AI in 6G systems at the PHY layer and above layers is discussed in \cite{stoica20196g}. The study in  \cite{6005345} covers a broad range of issues relating to taking advantage of THz frequency bands and provides an extensive review of the various radio-frequency (RF) hardware challenges that must be overcome for systems to operate in the THz bands. Collectively, the 6G vision developed by the studies mentioned above and by the current paper is summarized in Fig.~\ref{6GVision}.
\begin{figure*}[!t]
    \vspace{-10pt}
     \centering
     \includegraphics[width=12.2cm]{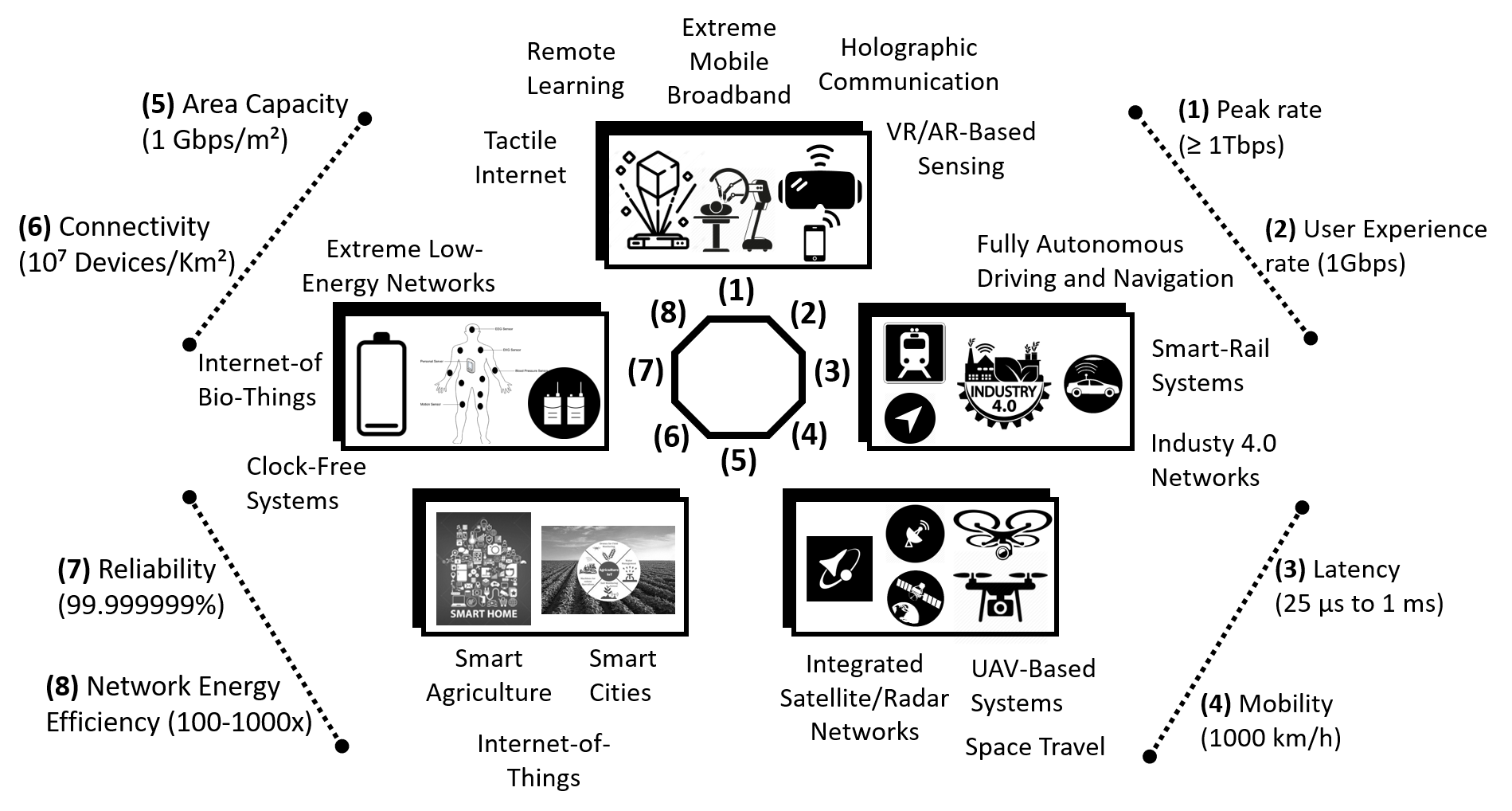}
     \vspace{-7pt}
     \caption{A vision for 6G systems and its underlaying use cases. Here we also summarize the key performance metrics which are of primary interest.} 
     \label{6GVision}
     \vspace{-14pt}
 \end{figure*}

\vspace{-12pt}
\subsection{Contributions of the Paper}
\label{ContributionsofthePaper}
\vspace{-1pt}
While the aforementioned and other papers cover important aspects of 6G systems, the aim of the current paper is to provide a \emph{holistic top-down} view of 6G system design. Starting from the the technical capabilities needed to support the 6G applications, we discuss the new spectrum bands which present an opportunity for 6G systems. While a lot of bandwidth is available in these new bands, how to utilize it effectively remains a key challenge, which we discuss in depth. For instance, frequency bands at 100 GHz and above present formidable challenges in the development of hardware and surrounding system components, limiting the application areas where all of the spectrum can be utilized. We discuss the deployment scenarios where 6G systems will most likely be used, as well as the technical challenges that must be overcome to realize the development of such systems. This includes new modulation methods, waveforms and coding techniques, multiple access techniques, antenna arrays, RF transceivers, real-time signal processing, as well as wave propagation aspects. We note that these are all substantial challenges in the way of systems that can be realized and deployed. Nevertheless, addressing these challenges at PHY layer is only a part of resolving the potential issues. Improvements in the network architecture are equally important. The present \emph{core network} design is influenced - and encumbered - by historical legacies. For example, the sub-millisecond latency required by many of the new services cannot be handled by the present transport network architecture. To this end, flattening or significant reduction of the architecture is necessary to comply with 6G use case requirements. The basic fabric of mobile internet - the Transmission Control Protocol/Internet Protocol  (TCP/IP) - is not able to guarantee quality-of-service (QOS) needed for many 6G applications, as it is in effect based on best effort services. These and many other aspects require a complete re-think of the network design, where the present transport networks will begin to disappear and be virtualized over existing fiber, as well as be isolated using modern software defined networking (SDN), and virtualization methodologies. At the same time, the core network functions will be packaged into a micro service architecture, and enabled on the fly.

\emph{All of these topics and more are covered in the sections below. For each aspect of 6G that is discussed in the paper, we present a detailed breakdown of the strengths and weaknesses of the presented concepts, technologies, or potential solutions. We differentiate what may be practically realizable, relative to what is theoretically possible. In doing so, we clearly highlight research challenges and unique opportunities for innovation created by these challenges.} To the best of our knowledge, a holistic contribution of this type is missing from the literature.

\vspace{-11pt}
\subsection{Organization of the Paper}
\label{OrganizationofthePaper}
The remainder of the paper is organized as follows. A vision for 6G, a discussion of \emph{seven} most prominent use cases to be supported by 6G, as well as their technical requirements are given in Sec.~\ref{vision}. A summary table of the KPIs and a comparison with 4G and 5G systems is also presented. This is followed by a discussion of the new frequency bands and deployment scenarios in Sec.~\ref{bands}. With the top-down approach, the fundamental changes in the core and transport networks supporting 6G applications is discussed in Sec.~\ref{core}. Complimenting this, a discussion of the new PHY techniques covering a wide range of topics such as waveforms, modulation methods, multiple antenna techniques, applications of AI and machine learning (ML) is contained in Sec.~\ref{modulation}. An overview of wave propagation characteristics of 6G systems for different applications and scenarios is given in Sec.~\ref{models}. The challenges in building radio transceivers and performing real-time signal processing for 6G, as well as solutions to overcome them are described in 
Sec.~\ref{transceiverdesign}. Finally, the conclusions are given in Sec.~\ref{conclusions}. A comprehensive bibliography is provided for the reader to delve deeper.

\vspace{-7pt}
\section{6G Use Cases and Technical Requirements}
\label{vision}
We now discuss the system requirements for 6G use cases. It is clear that the major applications and usage scenarios for 6G discussed above require instantaneous, extremely high speed wireless connectivity \cite{Li,Latva,Yastrebova}. The system requirements for Network 2030 have recently been published by the ITU-T in \cite{ITUTFGNET2030subgroupg1}\footnote{According to the ITU-T in \cite{ITUTFGNET2030subgroupg1}, \emph{system} requirements as denoted in our terminology are referred to as \emph{network} requirements. To avoid ambiguity with the \emph{network layer} of the OSI stack, we avoid the use of the term \emph{network requirements}.}. Here we review these, as well as requirements published in other sources quoted above. We categorize the requirements separately for each 6G use case in the subsections below.

\vspace{-12pt}
\subsection{Use Case 1: Holographic Communications}
\label{UseCase1HolographicCommunications}
As discussed earlier, holographic displays are the next evolution in multimedia experience delivering 3D images from one or multiple sources to one or multiple destinations, providing an immersive 3D experience for the end user. Interactive holographic capability in the network will require a combination of very high data rates and ultra low latency. The former arises because a hologram consists of multiple 3D images, while the latter is rooted in the fact that parallax is added so that the user can interact with the image, which also changes with the viewer's position. This is critical in providing an immersive 3D experience to the user \cite{khaled}. The key system requirements for this type of communication are:
\begin{enumerate}
\item \underline{\emph{Data rates}:} The data rates required depend on how the hologram is constructed, as well as on the display type and the numbers of images which need to be synchronized. Data compression techniques may reduce the data rates needed for the transmission of holograms, but even with compression, holograms will require massive bandwidths. These vary from tens of Mbps \cite{holoport} to 4.3 Tbps \cite{Li,8970173} for a human size hologram using image-based methods of generating holograms.
\item \underline{\emph{Latency}}: Truly immersive scenarios require ultra low latency, else the user feels simulator sickness \cite{8970173}. Nevertheless, if haptic capabilities are also added then sub-millisecond latency is required \cite{Atsushi,ITUTFGNET2030subgroupg1}. This is elaborated in Use Case 2 in the following subsection. 
\item \underline{\emph{Synchronization}}: There are many scenarios where synchronization needs to be adhered to in holographic communications. As different senses may get integrated, the different sensor feeds may be sent over different paths or flows, and will require synchronization, as well as coordinated delivery. When streams involve data from multiple sources, such as video, audio, and tactile, precise/stringent inter-stream synchronization is required ensuring timely arrival of the packets. Coordinated delivery of the flows need dependency objectives for time-based dependency, ordering dependency and QOS fate sharing. \emph{For all of this to happen, the network must have knowledge of the co-flows - something which is non trivial.} Another example is the case of a \emph{virtual orchestra}, whereby members of the orchestra are in different locations, and their movements must be coordinated such that it seems as if the music is emanating from the same stage.\footnote{While this is managed currently in 5G systems with 2D images, the complexity and challenges for a problem of such type with holographic communication is an order-of-magnitude greater.} Multiparty robotic communications via holograms is yet another example where communication  between a leader and a follower or between multiple robotic agents requires synchronization \cite{8322220}. 
\item \underline{\emph{Security}}: Requirements for this depend upon the application. If remote surgery is to be carried out, then the integrity and security of that application is absolutely vital, as any lapse could be life threatening. Coordinating the security of multiple co-flows is an additional challenge, as an attack on a single flow could compromise all other members of the flow. 
\item \underline{\emph{Resilience}}: At the system level, resilience is about minimizing packet loss, jitter and latency. At the service level, relevant quality-of-experience metrics are availability and reliability. For holographic communication services, an un-recovered failure event could pose a significant loss of value to operators. Therefore,  system (network) resilience is of paramount importance to maintain the high QOS needs for these services. 
\item \underline{\emph{Computation}}: There are significant real-time computational challenges at each step of hologram generation and reception. While compression can reduce the bandwidth needs, it will heavily influence the latency incurred. To this end, there is an important trade-off between higher level of compression, computation bandwidth and latency which needs to be optimized. A discussion on this is contained in \cite{8322220}.
\end{enumerate}
We note that there are significant challenges in the realization of holograms and multi-sense communications - especially for their wide spread adoption \cite{BLINDER1}. These challenges apply in all stages of the holographic video systems and range from signal generation to display. Current holographic displays are limited to head mounted displays (HMDs). To the best of our knowledge, there are no standards that specify how to supply data to a display. The recording of digital holograms is another challenge, as specialized optical setups may be required. Computer generated holograms are highly computation intensive in comparison with classical image rendering due to the many-to-many relationships between a source and hologram pixels. The large data rates required cannot take advantage of established compression techniques like joint photographic experts group (JPEG)/moving picture experts group (MPEG), since the statistical properties of holographic signals are much different from motion video. Though current HMDs only require on the order of 100 Mbps, they are more suitable for AR/VR applications, and offer limited 3D effects without accounting for several cues of the human visual system. Continued HMD use could lead to eye strain and nausea. As for using a mobile device to experience a hologram, there are additional graphics processing unit (GPU) and battery life limitations. The GPU performance of a mobile device is typically 1/40-th of an average personal computer GPU \cite{LEESAMSUNGTALKICC20}, requiring a significant improvement to meet the service requirements of holograms. The authors of \cite{BLINDER1} give a summary of the challenges that need to be tackled to pave the way for the realization of dynamic holographic content.

\vspace{-10pt}
\subsection{Use Case 2: Tactile and Haptic Internet Applications}
\label{UseCase2TactileandHapticInternetApplications}
There are many applications that fall in this category \cite{7403840}. Consider the following examples: 
\begin{itemize}
\item \emph{Robotic and Industrial Automation:} We are at the cusp of witnessing a revolution in manufacturing stimulated by networks that facilitate communications between humans, as well as between humans and machines in Cyber-Physical-Systems (CPS) \cite{7019732}. This so-called \emph{industry 4.0} vision is enabling a plethora of new applications \cite{Brettel2014HowVD}.\footnote{We note that the previous three industrial revolutions were triggered by water and steam - industry 1.0, mass production assembly lines and electrical energy - industry 2.0, as well as automated production using electronics and IT - industry 3.0.}
It requires communications between large connected systems without the need for human intervention. Remote industrial management is based on real-time management and control of industrial systems. Robotics will need real-time guaranteed control to avoid oscillatory movements. Advanced robotics
scenarios in manufacturing need a maximum latency
target in a communication link of 100 microseconds ($\mu$s), and round-trip reaction times of 1 millisecond (ms). Human operators can monitor the remote machines by VR or holographic-type communications, and are aided by tactile sensors, which could also involve actuation and control via kinesthetic feedback. 

\item{\emph{Autonomous Driving:}} Enabled by vehicle-to-vehicle (V2V) or vehicle-to-infrastructure communication (V2I) and coordination, autonomous driving can result in a large reduction of road accidents
and traffic jams. However, a latency in the order of a few ms will likely be needed for collision avoidance and remote driving. Thus, advanced driver assistance, platooning of vehicles, and fully automated driving, are the key application areas that 6G aims to support, and mature, with first components to be implemented in Third Generation Partnership Project (3GPP) Release 16 \cite{3GPPV2X1}; see also a list of use cases by the 5G Automotive Association (5GAA) in \cite{5GAA}. Yet, since no fully functional autonomous vehicles exist, further requirements and applications are sure to emerge over the next decade within this area. 

\item{\emph{Health Care:}} Tele-diagnosis, remote surgery and tele-rehabilitation are just some of the many potential applications in healthcare. We have already witnessed an early form of this during the ongoing COVID-19 pandemic, whereby a huge number of medical consultations are via video links. However, with the aid of advanced tele-diagnostic tools, medical expertise/consultation could be available anywhere and anytime regardless of the location of the patient and the medical practitioner. Remote and robotic surgery is an application where a surgeon gets real-time audio-visual feeds of the patient that is being operated upon in a remote location. The surgeon operates then using real-time visual feeds and haptic information transmitted to/from the robot; this is already happening in some instances, see e.g., \cite{4209212}. The tactile internet is at the core of such a collaboration. The technical requirements for haptic internet capability cannot be fully provided by current systems as discussed in  \cite{Antonakoglou}.
\end{itemize}
The key network requirements for these type of services are:\vspace{3pt}
\begin{enumerate}
\item\underline{\emph{Data rates}}: Data rates depend upon the application requirements \cite{8322220}: For example, a high definition 1080p video only needs 1-5 Mbps, 4K 360$^{\circ}$ video needs 15-25 Mbps \cite{shafi20175g}, whereas a hologram via point cloud techniques requires 0.5-2 Gbps, with large-sized holograms needing up to a few Tbps. For another application, such as autonomous driving, multiple sensors on next generation cars could result in an \emph{aggregate} data rate of 1 Gbps to be used for V2V and vehicle-to-everything (V2X) scenarios \cite{GONZALEZ1}. 
\item\underline{\emph{Latency}}: The human brain has different reaction times to various sensory inputs ranging from 1 to 100 ms \cite{Fettweis}. While it takes 10 ms to understand visual
information, and up to 100~ms to decode the audio signals, only 1~ms is required to
receive a tactile signal. Thus, the tactile internet requires  end-to-end latency on the order of 1 ms\cite{Fettweis}, and sub-ms latency may be required for instantaneous haptic feedback, otherwise conflicts between visual, and other sensory systems could cause cyber sickness to the tactile users \cite{7403840}. Robotics and other industrial machinery will also need sub-ms latencies. 
\item\underline{\emph{Synchronization}}:  Due to the fast reaction times of the human mind to tactile inputs, different such real-time inputs arising from different locations must be strictly synchronized. Similarly, as machine control might have fast reaction times, their inputs need to be tightly (sub-ms level) synchronized as well. 
\item\underline{\emph{Security}}: For all of the above applications (from robotics to autonomous cars), we envisage security to be at the forefront of the potential issues. This is since an attack/failure on/of particular system functionality could lead to life threatening situations. 
\item\underline{\emph{Reliability}}: Some applications, such as cooperative autonomous driving and industrial automation demand a level of reliability that wireless systems of today are not able to guarantee. Ultra reliable transmissions are  
assumed to have a success rate of ``five nines", i.e., 99.999\% \cite{6882630}. Industrial IOT systems could require even higher reliability, such as 99.99999\% \cite{8584994}, since loss of information could be catastrophic in some cases.
\item\underline{\emph{Prioritization}}: The network should be able to prioritize streams based on their criticality. Visual feeds may have many views with different priorities. 
\end{enumerate}

\vspace{-11pt}
\subsection{Use Case 3: Network and Computing Convergence}
\emph{Mobile edge compute (MEC)} will be deployed as part of 5G networks, yet this architecture will continue towards 6G networks. When a client requests a low latency service, the network may direct this to the nearest edge computing site. For computation-intensive applications, and due to the need for load balancing, a multiplicity of edge computing sites may be involved, but the computing resources must be utilized in a coordinated manner.\footnote{A more general form is {\em Augmented Information Services}, where computations are performed on data streams that are transmitted in a multi-hop fashion from a transmitter to the receiver, and the computations can be performed at intermediate nodes - see \cite{8500763} for further details.} Augmented reality/virtual reality (AR/VR) rendering, autonomous driving and holographic type communications are all candidates for edge cloud coordination. The key network requirements for this are: computing awareness of the constituent edge facilities, joint network and computing resource scheduling (centralized or distributed), flexible addressing (every network node can become a resource provider), fast routing and re-routing (traffic should be able to route or re-route in response to load conditions). Fig.~\ref{CloudCoordinationRAN} demonstrates this vision via edge-to-edge coordination across local edge clouds of different network and service types, as well as edge coordination with the core cloud architecture. 
 \begin{figure}[!t]
 \vspace{-7pt}
     \centering
     \hspace{-10pt}
     \vspace{-2pt}
     \includegraphics[width=8.3cm]{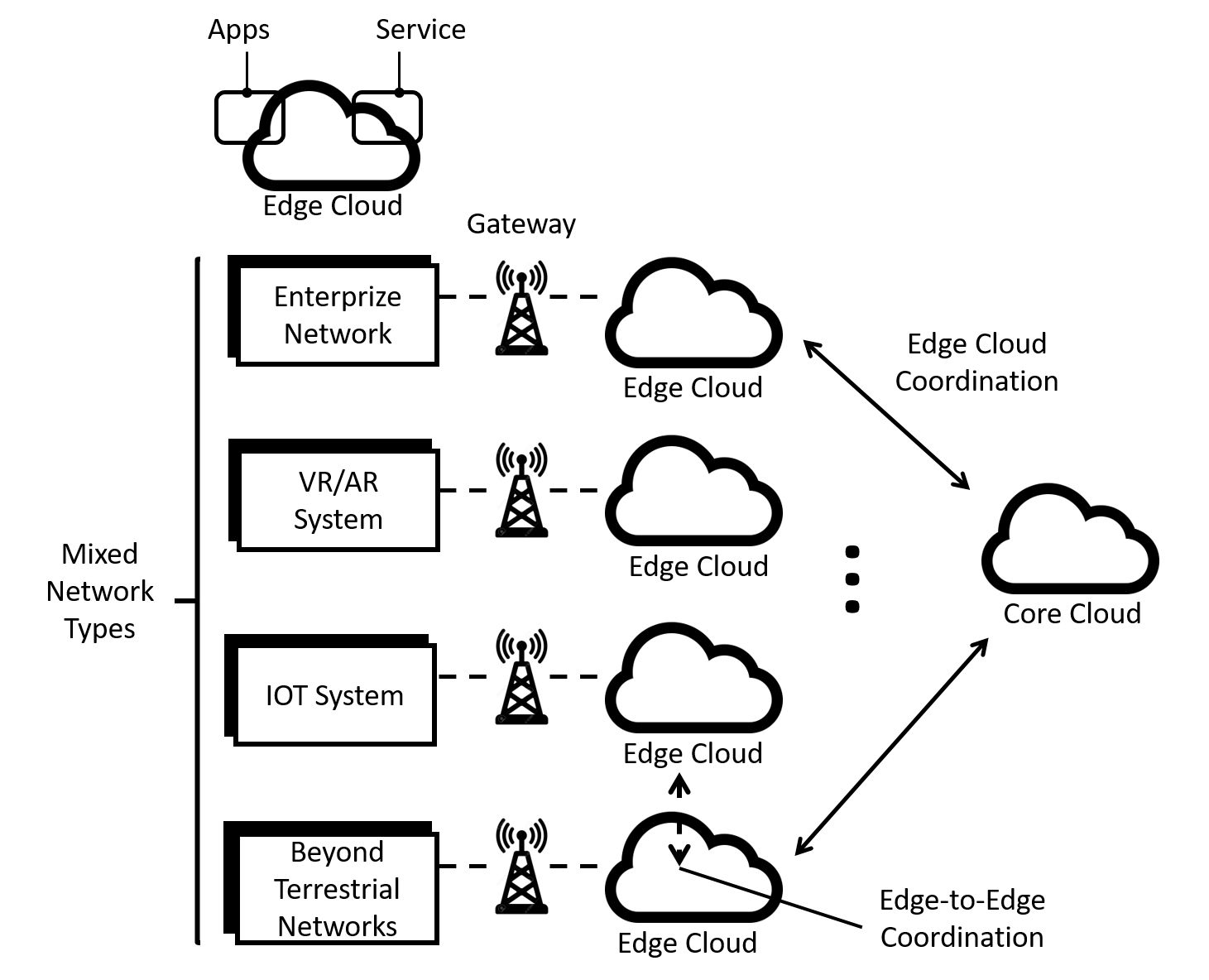}
     \caption{Cloud coordination between local edges driven by different network types and services, as well as across the local edge cloud and core cloud. The figure is inspired from the discussions in \cite{ITUTFGNET2030subgroupg1}. }
     \label{CloudCoordinationRAN}
 \end{figure}
\begin{figure}[!t]
     \centering
     \hspace{2pt}
     \includegraphics[width=8.4cm]{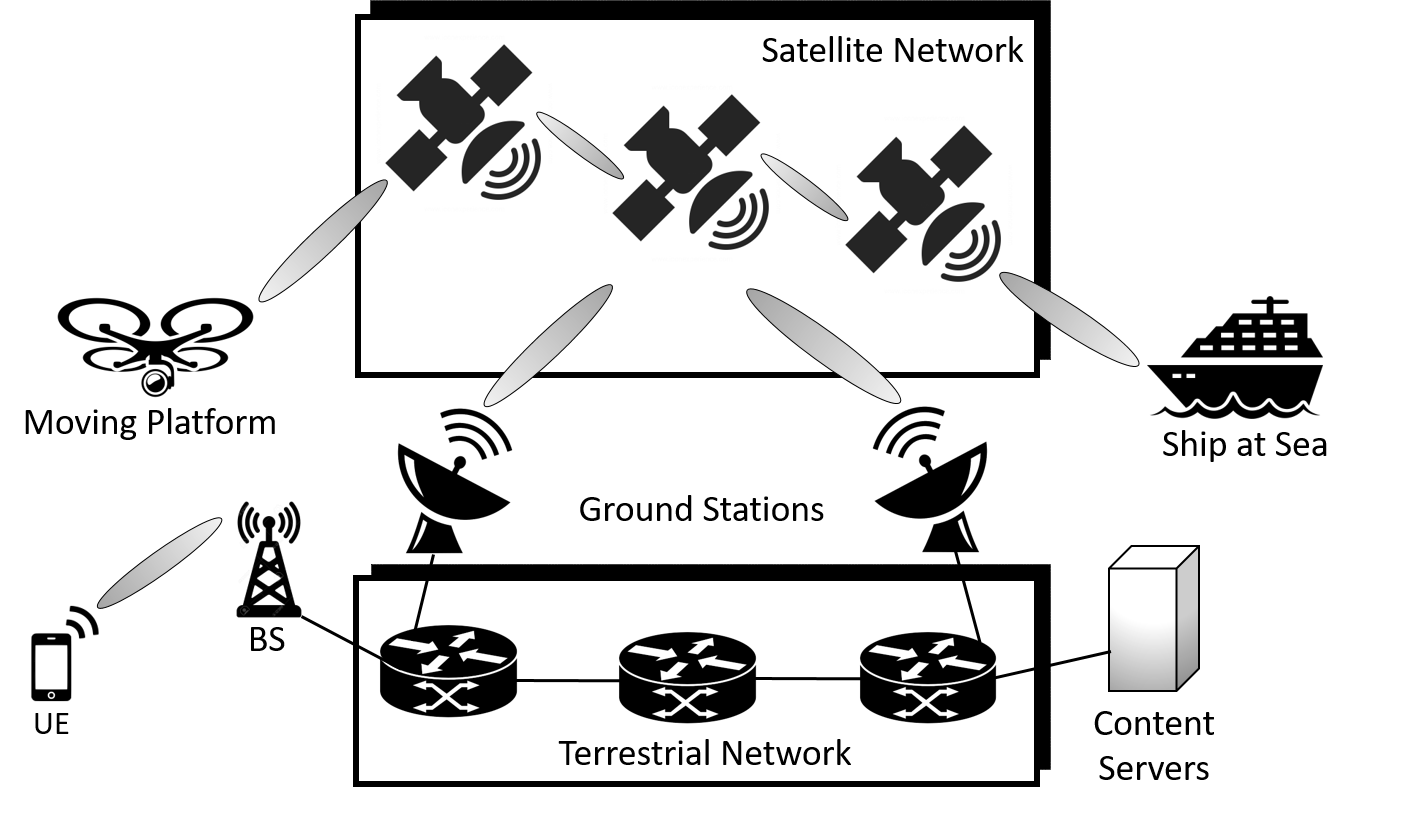}
     \caption{Space integrated terrestrial networks incooperating multiple moving platforms in a unified framework. The figure is inspired from \cite{ITUTFGNET2030subgroupg1}.}
     \label{SpaceTerrestrialNetworkIntegration}
     \vspace{-13pt}
 \end{figure}
 
\vspace{-5pt}
\subsection{Use Case 4: Extremely High Rate Information Showers} 
\vspace{1pt}
Access points in metro stations, shopping malls, and other public places may provide information shower kiosks \cite{Petrov}. The data rates for these information shower kiosks could be up to 1 Tbps. The kiosks will provide fibre-like speeds. They could also act as the backhaul needs of millimeter-wave (mmWave) small cells. Co-existence with contemporaneous cellular services as well as security seems to be the major issue requiring further attention in this direction. 
\vspace{-10pt}
\subsection{Use Case 5: Connectivity for Everything} 
This use case can be extended to various scenarios that include real-time monitoring of buildings, cities, environment, cars and transportation, roads, critical infrastructure, water and power etc. Besides these use cases, internet of bio-things through smart wearable devices, intra-body communications achieved via implanted sensors will drive the need of connectivity much \emph{beyond mMTC}. The key network requirements for these use cases are: Large aggregated data rates due to vast amounts of sensory data, high security and privacy in particular when medical data is being transmitted, as well as possibly low latency when a fast intervention (e.g., heart attack) is required. As yet no systems or models exist to assess these data needs.

\vspace{-10pt}
\subsection{Use Case 6: Chip-to-Chip Communications} 
While on-chip, inter-chip, and inter-board communications nowadays are done through wired connections, those links are becoming bottlenecks when the data rates are exceeding 
100-1000 Gbps. There have thus been proposals to employ either optical or THz wireless connections to replace wired links. The development of such ``nanonetworks" constitutes another promising area for 6G. Important criteria for such networks - besides the data rate - is the energy efficiency (which needs to incorporate possible required receiver processing), reliability, as well as latency. Specific KPIs for nanonetworks depend on chip implementations and applications, which will become clearer as they are developed over the next decade.  

\vspace{-10pt}
\subsection{Use Case 7: Space-Terrestrial Integrated Networks}
\vspace{-3pt}
This use case presents a scenario that is based on internet access via the seamless integration of terrestrial and space networks. The idea of providing internet from space using large constellations of LEO satellites has re-gained popularity in the last years (previous attempts such as the Iridium project in the late 1990s had failed). The study in \cite{LEOs} compares Telesat's, OneWeb's, and SpaceX's satellite systems. The key benefits of these are: Ubiquitous internet access on a global scale including on moving platforms (aeroplanes, ships, etc.), enriched internet paths due to the border gateway protocols across domains relative the terrestrial internet, and ubiquitous edge caching as well as computing. The mobile devices for these integrated systems will be able to have satellite access without relying on ground base infrastructures. The key network requirements for this capability are: (1) Flexible addressing and routing; with thousands of LEO satellites there are new challenges for the terrestrial internet infrastructure to interact with the satellites. (2) Satellite bandwidth capability: The inter-satellite links and terrestrial internet infrastructure in some domains could be a bottleneck for satellite capacity. (3) Admission control by satellites: When a satellite directly acts as an access point, this requires each satellite to have knowledge about the traffic load in the space network to make admission control decisions. (4) Edge computing and storage: The realization of edge computing and storage will incur challenges on the satellite due to on-board limitations. Latency will also be a challenge as the physical distance between the satellite and end node will set a limit on the minimum delay introduced by the link. An example realization of space-terrestrial integrated networks is depicted in Fig.~\ref{SpaceTerrestrialNetworkIntegration}, where multiple services communicating to the satellite network and terrestrial networks are shown to seamlessly co-exist. 

Collectively, in view of the above, the key requirements for 6G systems may be summarized (in the style of corresponding requirements for 5G systems) as \cite{ITURminimum,Latva}:
\begin{itemize}
\item \emph{Peak data rate}: $\geq$1 Tbps catering to holographic communication, tactile internet applications and extremely high rate information showers. This at least 50$\times$ larger than that of 5G systems.
\item \emph{User experience data rate}: At least be 10 $\times$ that of the corresponding value of 5G.  
\item \emph{User plane latency}: This is application dependent, yet its minimum should be a factor 40$\times$ better than in 5G. 
\item \emph{Mobility:} It is expected that 6G systems will support mobility of upto 1000 km/h to include mobility values encountered in dual-engine commercial aeroplanes.
\item \emph{Connection density per-km$^{2}$:} Given the desire for 6G systems to support an internet-of-everything, the connection density could be 10$\times$ that of 5G. 
\end{itemize}
\vspace{-5pt}
The above capabilities and more are summarized in  Tab.~\ref{table1}, relative to the corresponding values in 5G and 4G systems. Realizations of the technical capabilities as discussed in this are significant challenges which must be overcome.
\begin{table*}[!t]
\centering
\scalebox{0.77}{
\begin{tabular}{cccc}
\toprule 
\textbf{KPI} & \textbf{4G} & \textbf{5G} & \textbf{6G}
\tabularnewline
\midrule
\midrule 
\textbf{Operating Bandwidth} & Up to 400 MHz & Up to 400 MHz for sub-6 GHz bands & Up to 400 MHz for sub-6 GHz bands\tabularnewline
& (band dependent) & (band dependent) & Up to 3.25 GHz for mmWave bands\tabularnewline
& & Up to 3.25 GHz for mmWave bands & Indicative value: 10-100 GHz for THz bands \tabularnewline\hline
\textbf{Carrier Bandwidth} & 20 MHz & 400 MHz & To be defined  \tabularnewline\hline
\textbf{Peak Data Rate} & 300 Mbps with 4$\times$4 arrays & 20 Gbps & $\geq$1 Tbps\tabularnewline 
& 150 Mbps with 2$\times$2 antenna arrays& &(Holographic, VR/AR, and tactile applications)\tabularnewline\hline
\textbf{User Experience Rate}& 10 Mbps (shared over UEs) & 100 Mbps & 1 Gbps\tabularnewline\hline
\textbf{Average Spectral Efficiency} &  25 Mbps with 2$\times$2 antenna arrays & 7.8 bps/Hz (DL) and 5.4 bps/Hz (UL) & 1$\times$ that of 5G\tabularnewline
& 40-45 Mbps with 4$\times$4 antenna arrays & & \tabularnewline\hline
\textbf{Connection Density} & N/A & $10^6$ devices/km$^2$ & $10^7$ devices/km$^2$
\tabularnewline\hline
\textbf{User Plane Latency} & 50 ms & 4 ms (eMBB) and 1 ms (uRLLC) & 25 $\mu$s to 1 ms\tabularnewline
& & & (Holographic, VR/AR and tactile applications) \tabularnewline\hline
\textbf{Control Plane Latency} & 50 ms & 20 ms & 20 ms\tabularnewline\hline
 \textbf{Mobility} & 350 km/h & 500 km/h & 1000 km/h\tabularnewline
 & & & Handling multiple moving platforms\tabularnewline\hline
\textbf{Mobility Interruption Time} & N/A & 0 ms (uRLLC) & 0 ms\tabularnewline
& & & (Holographic, VR/AR and tactile applications)\tabularnewline\bottomrule
\end{tabular}}
\vspace{7pt}
\caption{Technical performance requirements of 6G systems and a comparison of the 6G KPIs relative to those for 5G and 4G systems. }
\vspace{-14pt}
\label{table1}
\end{table*}

\vspace{-10pt}
\section{New Frequency Bands and Deployments}
\label{bands}
\vspace{-2pt}
\subsection{New Frequency Bands for 6G}
\vspace{-3pt}
Traditionally, new generations of wireless systems have exploited new spectrum in order to satisfy the increased demands for data rates. 5G systems are characterized to a significant degree by the use of the mmWave spectrum complimented by large antenna arrays. A further expansion to higher frequencies for 6G seems almost unavoidable. \emph{However, we note that not all 6G services will be suitable to be offered in the new bands. The existing bands for 4G and 5G will continue and may be re-farmed for 6G.} In this spirit, the spectrum from 100~GHz to 1~THz is being considered as a candidate for 6G systems. Within this band, particular sub-bands have very high absorption (see Sec.~\ref{models} for a discussion of the physical reasons) and are thus ill suited for communication over more than a few meters. The \emph{spectrum windows} with lower absorption losses shown in Fig.~\ref{AbspPeaks} still represent a substantial amount of aggregated bandwidth \cite{akyildizdistance,akyildiz,jornet2011channel,chonghan}. Nevertheless, this spectrum is also used by various existing services. Consequently, all of it will likely not be made available by frequency regulators, and also not allocated in a contiguous manner. In particular, over the range of 141.8 GHz to 275 GHz, there are various blocks containing existing services that have co-primary allocation status by the ITU. These services include fixed, mobile, radio astronomy, earth exploration satellite service (EESS) passive, space research passive, inter-satellite, radio navigation, radio navigation satellite, and mobile satellite systems. Amongst the above, the passive services are much more sensitive to interference, and their protection will require guard bands, limits on out-of-band emissions as well as in-band transmit power, restrictions on terrestrial beams (by controlling the power flux densities), and side lobes pointing upwards. All of these aspects are critical for the co-existence of terrestrial systems with space-based networks. The next World Radio Conference (WRC) in 2023 will consider allocation of 231.5 GHz to 252 GHz to EESS passive systems. Parts of the spectrum beyond 257 GHz are also allocated to various other passive services. The authors in \cite{6005345} expound the difficulties of co-existence between radio astronomy and wireless services in THz bands. Despite all of the above, the amount of spectrum available represents a unique opportunity for 6G systems. 

The use of the above-mentioned frequency windows is dependent upon a specific use case; naturally, not all the windows will be suitable for all use cases. The first window of interest will be the one marked as W1 in Fig.~\ref{AbspPeaks} covering the frequency range from 140-350 GHz. This band is typically referred to as the sub-THz band, even though strictly speaking ``high mmWaves" might be the more appropriate nomenclature. The two key advantages of this band are: 1) The existence of many tens of GHz of bandwidth that are currently lying unused; and 2) The ability to develop ultra massive multiple-input multiple-output (MIMO) antenna arrays within a reasonable form factor. The use of spectrum in higher windows is accompanied by a higher absorption loss. Though Fig.~\ref{AbspPeaks} is shown up to 1 THz, one can go even higher in frequency up to 10 THz \cite{4263933,6005345} at the expense of beyond formidable hardware realization challenges, so that this use seems further away. From this point onward, a move to even higher frequency bands brings us to some familiar territory, namely that of free-space optical (including infrared) links, either through the use of laser diodes, or light emitting diodes (LEDs) commonly assumed for visible light communications (VLC). Both of these approaches have been explored for a number of years, but it is only recently that an integration into cellular and other wireless systems seems to increasingly become a realistic option. 
\vspace{-10pt}
\subsection{6G Deployment Scenarios}
\vspace{-1pt}
Besides the exploration and the use of new frequency ranges, an investigation into new deployments is necessary. While some applications of 5G will also continue to be deployed in the existing 5G bands, which over time may be refarmed to 6G, we identify possible new deployment scenarios primarily motivated by the previously unexplored THz bands. We note that there will naturally be many applications such as Connectivity for Everything (see Use Case 5 in Sec.~\ref{vision}) which will be in existing the sub 1 GHz band where a lot of the IOT deployments are happening. Another example is cellular V2X communication intended for autonomous driving, which will use a combination of microwave and mmWave bands \cite{GONZALEZ1}. 
\begin{figure}[!t]
\vspace{-15pt}
    \centering
    \hspace{-3pt}
    \includegraphics[width=8.4cm]{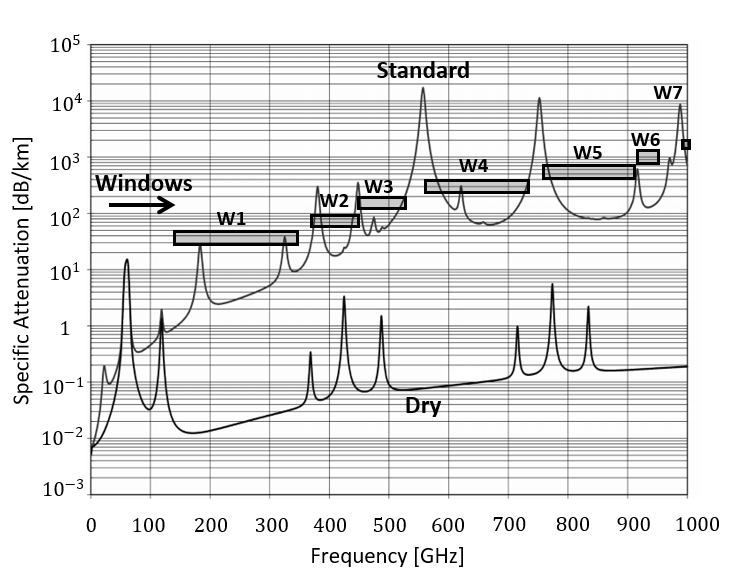}
    \vspace{-8pt}
    \caption{Average atmospheric absorption loss vs. carrier frequency up to 1000 GHz. The two curves denote the standard, i.e., sea-level attenuation and dry air attenuation, where various peaks and troughs are observed for oxygen and water sensitive regions. The figure is reproduced from  \cite{ITURP676}.}
    \label{AbspPeaks}
    \vspace{-14pt}
\end{figure}

\subsubsection{\underline{Hot Spot Deployments}} 
This is a conventional application whereby extremely high data rate systems (such as those described in Use Case 4) could be deployed indoors or outdoors. MmWave and THz systems, e.g., in the window W1, would be well suited for such scenarios. 
However, ubiquitous deployments will be uneconomical as coverage radius in outdoor environments is limited to about $100$ m, and even less in indoor environments - this follows from both free-space pathloss (even with reasonable-sized antenna arrays) and molecular absorption \cite{akyildizdistance,Jorneymodel}, see Tab.~\ref{table:2}. If more bandwidth is needed, we can aggregate more windows, though this might further shorten the feasible transmission distance. The authors of \cite{akyildizdistance} proposes a bandwidth vs. distance scheduling,  whereby more bandwidth is available for a lower transmission distance (say all the windows), and this progressively reduces to W1 for large distances. However, all of the link budgets only consider free-space pathloss. Further consideration of obstructing objects, scattering, and other effects need to be taken into account for realistic deployment planning.
\begin{table*}[!t]
\centering
\scalebox{0.71}{
\begin{tabular}{ccccccc}
\toprule 
\textbf{Window \#} & \boldsymbol{$f_{c}$} \textbf{[THz]} & \boldsymbol{$B_\textrm{3dB}$} \textbf{[GHz]} & \textbf{Loss at 10 mm [dB]}&\textbf{Loss at 1 m [dB]} &\textbf{Loss at 100 m [dB]}&\textbf{Absorption Loss [dB/Km]}
\tabularnewline
\midrule
\midrule 
\textbf{W1} &0.245 & 210  &60.18  &80.18  &120.18 &3  \tabularnewline\hline 
\textbf{W2} &0.41 & 65.61 &64.65  &84.65  &124.65 &20    \tabularnewline\hline 
\textbf{W3} &0.49 & 86.21&66.2  &86.2  &126.2 &40   \tabularnewline\hline
\textbf{W4} &0.66  &152.59  &68.79  &88.79  &128.79 &60  
\tabularnewline\hline
\textbf{W5} &0.84  &141.91  &70.88  &90.88  &130.88 &80  
\tabularnewline\hline
\textbf{W6} &0.94  &47.3   &71.86  &91.86  &131.86 & 150 \tabularnewline\hline
 \textbf{W7} &1.03  &57.98  &72.65  &92.65  &132.65 & -- \tabularnewline\bottomrule
\end{tabular}}
\vspace{8pt}
\caption{Operating Windows in THz Bands. Free space loss is calculated at the center frequency of each window. Absorption loss is obtained from Fig.~\ref{AbspPeaks} for ``standard" atmospheric conditions.}
\vspace{-20pt}
\label{table:2}
\end{table*}
\subsubsection{\underline{Industrial Networks}}
While 5G was innovative in introducing the concept of industry 4.0, we anticipate that 6G will take significant strides in transforming the manufacturing and production processes. The maturity of industrial networks will depend on a successful adoption of current and future radio access technologies to the key industry 4.0 and beyond use cases. Industrial networks are envisaged to be privatized, focusing on extreme reliability and ultra low latency. The key deployment use cases are: 1) Communication between sensors and robots; 2) Communications across multiple robots for coordination of tasks; and 3) Communication between human factory operators and robots. Currently, in order to achieve the requirements for ultra high reliability, the majority of the commercial deployments are taking place between 3.4-3.8 GHz, where the propagation channel is relatively rich in terms of diffraction efficiency \cite{3GPP1,5GSMART1}. Yet machines with massive connectivity in the 6G era will also demand high data rates along side real-time control and AI to be able to transmit and process high-definition visual data, enabling digital twins of machines and operations, as well as remote troubleshooting. To this end, we foresee the use of mmWave frequencies in addition to bands below 6 GHz for industrial networks over the next decade. Preliminary studies such as the one in \cite{ALSAADEH1} are demonstrating possibilities and challenges of integrating mmWave frequencies within industry 4.0 scenarios.
\subsubsection{\underline{Wireless Personal Area Networks (WPANs)}}Another area of deployment is WPANs and wireless local area networks (WLANs). These could be in between a laptop and an access point, an information kiosk and a receiver\cite{Hojinsong}, between AR/VR wearables and a modem  or between the ``infostations" proposed in \cite{goodman1997infostations}. These are very short links perhaps less than 0.5~m to 1~m for WPANs, and up to 30 m for WLANs. All windows may be suitable for this application provided the link budget can meet the path loss when the higher windows are used and where appropriate implementation technologies exist.

\subsubsection{\underline{Autonomous Vehicles and Smart Railway Networks}}6G could be used for information sharing between autonomous vehicles and V2I \cite{Ota}. However, there are doubts if the complicated traffic conditions and short distances due to range limitation discussed earlier will make the THz bands suitable for this application. Furthermore, high-speed adaptive links between antennas on train rooftops and infrastructure can be used for transmission of both safety-critical information and aggregate passenger data \cite{guan2017milllimeter,Ai_et_al_2020_ProcIEEE}. Such extremely-high rate links are well suited for THz, yet the high mobility creates strong sensitivity to beamforming errors and possible issues with the Doppler spread. While the speed of mordern high-speed trains is almost constant, and thus beams can be steered into the right direction based on prediction, the required beamforming gain (and associated narrow beamwidth) make the system sensitive to even small deviations from the predictions \cite{Kim_and_Molisch_2013}. 
Furthermore, high-frequency systems can also be used for access between UEs and antennas in the cabins that aggregate the passenger data, similar to a (moving) hotspot. 

Keeping in mind the emerging 6G use cases, technical requirements, new frequency bands and key deployment scenarios, in the following section, we discuss the changes required to the design of 6G radio and core network architectures.

\vspace{-3pt}
\section{6G Radio and Core Network Architectures: Design Principles and Fundamental Changes}
\label{core}
In order to cater for the next generation use cases, 6G will consolidate many of the disruptive approaches introduced by 5G. Notably, the 5G standardization efforts have provided the ground work to enable flexible topologies to be deployed, breaking the traditional centralized hierarchy that exists today. KPIs such as latency, can be tailored to use cases thanks to innovative features like network slicing, control/user plane separation and MEC. The service-driven architecture with atomized and largely API'ed software components allows already today for a much more open innovation community, thus helping to accelerate the pace of deployment. 6G will however introduce entirely novel paradigms. These will be novel features and capabilities; a novel thinking towards the underlying transport architecture infrastructure; and novel philosophies around the entire design process, which will hopefully accelerate design and deployment even further. These are discussed in the following.

\vspace{-9pt}
\subsection{6G Network Design Principles}
\vspace{-1pt}
Concerning novel protocol and architecture approaches, the following will be of notable importance:
\begin{enumerate}
\item {\underline{\em Super-Convergence}:} Non 3GPP-native wired and radio systems will form an integral part of the 6G eco-system. In fact, many of the more disruptive changes discussed below will not be possible without an easier and more scalable convergence between different technology families. Emphasis will be on mutual or 3GPP-driven security and authentication of said converged network segments. As such, wireline and wireless technologies like WiFi, WiGig, Bluetooth and others will natively complement 6G with the strong security and authentication methods of 3GPP used to secure the consolidated network. It will greatly aid with traffic balancing due to the ability to onboard  and offload traffic between networks of different loads; it will support resilience since traffic delivery can be hedged between different technology families.

\item {\underline{\em Non-IP Based Networking Protocols}:} Internet protocol version 6 (IPv6) is now decades old with calls for standardization of entirely novel networking protocols growing. Indeed, the body of research on protocols beyond IP is rich and several solutions are currently being investigated by European Telecommunications Standards Institute (ETSI)'s Next Generation Protocols (NGP) Working Group as possible candidates for such a disruptive approach. With more than 50\% of networking traffic originating in or terminating at the wireless edge, a solution which caters the wireless sector is fully justified.

\item {\underline{\em Information-Centric \& Intent-Based  Networks (ICN)}:} Related to above NGP, ICNs are an active research area in the internet research task force (IRTF) and internet engineering task force (IETF), and constitute a paradigm shift from networking as we know it today (i.e., TCP/IP based)~\cite{Fang}. ICN is a step toward the separation of content and its location identifier. Rather than IP addressing, content is addressed using an abstract naming convention. Different proposals exist today for the protocol realisation of ICN. It was considered in the ITU-T Focus Group (FG) on IMT-2020~\cite{FG2020} as a candidate for 5G. In fact, several proposals already exist to carry ICN traffic tunnelled through the mobile network but such an approach defies the transparent and flat Internet topologies. A new ITU-T FG has been established to guide the requirements for the network of 2030~\cite{FG2030}.
Furthermore, to bridge latest developments in networking design and operational management, intent-based networking as well as intent-based service design have emerged. It is a lifecycle management approach for networking infrastructure, which will be central to 6G. It will require higher-level business and service policies to be taken into account; a thus resulting system configuration leveraging on the end-to-end softwarized infrastructure; a continuous monitoring of the network and service state; and a real-time optimization process able to adapt to any changes in network/service state and thus ensuring that the intent is met. 

\item {\underline{\em 360-Cybersecurity \& Privacy-by-Engineering-Design}:} While security has been taken very seriously in 5G from a protocol and architecture point of view, the underlying embedded code which embodies and executes the various system components has never been part of the standardization efforts. Most security vulnerabilities however have been due to poorly written code. Thus, future efforts will not only focus on a secure end-to-end solution but will also encompass a top (architecture, protocols) - down (embedded software) approach which we refer to as 360-cybersecurity approach. Furthermore, whilst security-by-design is now a well understood design approach, privacy is still being solved at "consent" level. Privacy-by-Engineering-design will ensure that mechanisms are natively built into the protocols and architecture which would e.g. prevent the forwarding of packets / information if not certified to be privacy-vetted. 
For instance, a security camera will only be allowed to stream the video footage if certain privacy requirements are fulfilled at networking level and possibly contextual level, i.e. understanding who is in the picture and what privacy settings they have enabled.  

\item {\underline{\em Future-Proofing Emerging Technologies}:} 
A large swath of novel technologies and features is constantly appearing, the introduction of which into the telco architecture often takes decades. Examples of such technologies today are quantum, distributed ledger technologies (DLT) and AI. Tomorrow, another set of technologies will appear. All these ought to be embedded quicker and more efficiently which is why 6G needs to cater for mechanisms allowing not-yet-invented technologies to be embedded into the overall functional architecture. The subsequent subsection lays out some possible approaches to achieving this. Here, some more details on the specific technology opportunities of quantum, DLT and AI: The exciting features of quantum is that it can be used to make the 6G infrastructure tamper-proof. It can be used for cryptographic key exchanges and thus enabling a much more secure infrastructure. Furthermore, quantum computing enables a NP-hard optimization problems to be solved in linear time, thus allowing network optimization problems solved and executed in much quicker (if not real) time. 

DLTs enable data provenance in that data, transactions, contracts, etc, are stored and distributed in an immutable way. This proves useful in a large multi-party system with little or no trust between the involved parties. Whilst {DLTs rise to fame} in the financial world with the emergence of Bitcoin, the same industry dynamic applies to telecoms where different suppliers feed into the vendor eco-system, vendors into operators and operators serve consumers. DLTs allow for a much more efficient execution of all these complex relationships. For instance, a vendor feature approved by one operator with the approval stored on a given DLT, should make other operators trust the feature without the need for lengthy procurement processes. Another example is where consumers can create their own market place to trade data plans, or other assets as part of the telco subscriber plan.

Finally, AI has been used within telecoms for years, but mainly to optimize consumer facing issues, such as churn, or network related issues, such as the optimal base station (BS) antenna array tilt combined with the optimal transmission power policies. However, with the emergence of distributed and more atomized networks, novel forms of AI will be needed which can be executed in a distributed fashion. Furthermore, consumer-facing decisions will need to be explained thus calling for Explainable AI (xAI) concepts which are able to satisfy stringent regulatory requirements.
\end{enumerate}

\vspace{-10pt}
\subsection{Opportunities for Fundamental Change}
The underlying infrastructure, including the transport networks, will need to undergo substantial changes as the amount of traffic to be carried in 6G networks will be  orders-of-magnitude larger than what we will see in the next years with 5G networks. We expect the following fundamental changes: 
\begin{enumerate}

\item {\underline{\em Removal/Reduction of the Transport Network}:} Unknown to many, the transport (and attached core network functionalities) is in fact a \emph{legacy} artefact; we do have it in 5G because we had in 4G, and we have it in 4G because we had in 3G, and likewise 2G, and the reason it was introduced in 2G is because back then the internet was not able to provide the required QOS. However, today the transport fiber infrastructure is really well developed and there is no reason for operators to maintain their own private ``local area network (LAN) at national scale". A complete rethink may thus give the opportunity for the cellular community to solely focus on the wireless edge (air interface + radio access network + control plane to support all); and simply use a sliced Internet fiber infrastructure to carry the cellular traffic. Whilst it requires some policy and operational changes, the technologies to support such a modus operandi are there. 

\item {\underline{\em Flattened Compute-Storage-Transport}:} A flattened transport-storage-compute paradigm will be enabled by a powerful 6G air interface and a complete re-think of the core and transport networks as suggested above. A possible scenario is where transport is virtualized over existing fiber but isolated using modern SDN and virtualization methodologies. At the same time, the Core Network functions are packaged into a microservice architecture and enabled on the fly using containers or server-less compute architectures. To underpin novel gaming applications, we will also see a clearer split between central processing unit (CPU) and graphics processing unit (GPU) instructions sets, allowing each to be virtualized separately; for instance, the GPU instructions are handled locally on the phone whilst the CPU instructions executed on a nearby virtual MEC.

\item {\underline{\em Native Open Source Support}:} For economic and security reasons but also reasons related to quicker innovation cycles and thus quicker time-to-market, open source will be an ever-growing constituent of a 6G eco-system. This is corroborated by recent announcement of tier-1 operators going to use open source not only for their core network but also parts of the radio access network. This presents an exciting opportunity for the entire communications and computer science community, as features can be contributed at scale.

Furthermore, not only open source (input) but also open data (output) will be instrumental in unlocking the potential of 6G. Notably, many if not most design and operational decisions in 6G will be taken by some form of algorithms. Said algorithms need to be trained which requires huge amount of data. The telco eco-system has been historically conservative in opening up operational data; such as the amount and type of traffic carried over various  segments of the control and data planes. Automated mechanisms will need to be created in 6G which allow access to important data, whilst not compromising security of the network nor the privacy of the customers. 

\item {\underline{\em AI-Native Design Enabling Human-Machine Teaming}:} Machine Learning and AI have been part of 3GPP ever since the introduction of self-organizing networking (SON) in Release 8. However, the degrees of freedom, the high dynamics, the high disaggregation of 6G networks as well as more stringent policies will almost certainly require a complete rethink of how AI is embedded into the telco eco-system. 

6G is an exciting challenge for the AI community as there is no global technology eco-system which has such stringent design requirements on spatial distribution, temporal low-latency and high data volumes. Emerging paradigms, such as distributed AI, novel forms of transfer learning as well as ensemble techniques, need to natively fit the overall telecom architecture. 

Importantly, consumer-facing decisions taken by AI need to be compliant with various consumer-facing policies around the world, such as Article 13 in Europe's general data protection regulation (GDPR). This requires the disclosure of any "meaningful information about the logic involved, as well as the significance and the envisaged consequences of such processing for the data subject". 

As a result, novel paradigms such as \emph{Explainable AI (xAI)} will need to natively sit within 6G. This is because traditional AI based on deep learning operates like a ``black box" where even the design team cannot explain why the AI arrived at a specific decision. xAI is a set of methods and techniques allowing the results of the solution to be understood by humans. It is not only vital in the face of emerging regulation, but also improves the user experience by helping end users trust that the AI is making good decisions. Different xAI technology families are emerging today, with the most promising being based on Planning~\cite{Cashmore_icapsxai2019}.

Furthermore, AI will be used in the design process and not only operationally. We may not see it with 6G, but future networks will be designed by AI. We envisage a future where advanced AI/ML is able to scrape telco-related innovation from the Internet, translate it into code, self-validate that code, implement it into a softarized infrastructure, test it on beta users and roll it out globally, all in a few minutes rather than decades. It could potentially be the underpinning technology for a next generation industry platform, industry 5.0.

All of above will underpin novel design and operational paradigms leading to an unprecedented human-machine teaming to leverage on the strength of both.

\item {\underline{\em Human-Centric Networks}:} The telco ecosystem has evolved from an initially cell-centric architecture with designs in 2G and 3G  driven by cell coverage, and thus the base station placements. Today's 4G and 5G device-centric architecture is driven by capacity which is in turn directly linked to the amount of high-quality links the terminal (i.e. smartphone or a fixed wireless access modem) sees at any given time. These designs are very static and do not allow us to address important societal use cases, see Sec. II, where UEs of multiple users can simultaneously share radio resources but each having different KPI and QOS requirements. 6G has the opportunity to be human-centric in that it is societally aware and technologically adaptable, so that important societal needs or Black Swan events can be dealt with more efficiently and effectively. This fundamental change is vital as today's networking infrastructures have become too fragmented and heterogeneous to meaningfully support societal challenges. Examples of these shortcomings were laid bare with the ongoing COVID-19 crisis: a massive shift of networking resources from corporate premises to private homes was needed but unattainable in some cases; privacy concerns over tracing apps emerged but could not be dispelled since privacy was not fundamentally embedded into the infrastructure but rather provided through T\&C's; a significant increase in security breaches was reported by various agencies around the world. Additionally, the telco eco-system needs to communicate the impact of new technologies on health and well-being. This is because each new generation is being greeted with dooms-news which is not helpful to consumers nor the industry. 6G has the potential to revert this by spending considerable time analysing the impact of the frequency bands to be used onto human health and well-being, with findings well communicated. 6G will have a profound impact onto overall innovation cycle and the skills landscape of telecoms, providing a phenomenal opportunity for growth. This is illustrated through the high-level architecture in Fig.~\ref{fig:Architecture2} with the challenges and opportunities summarized in Tab.~\ref{6GDesignChallenges}.
\end{enumerate}
 
Let us examine the specific design use case of providing an extremely low-latency connectivity between two end points (UEs) pertaining to two different network operators. To this end, we discuss the transition from the current 5G network architecture to a possible 6G architecture. In 5G, low-latencies can be provided so long as the end points form a ``local area network", i.e. they belong to the same or physically close set of distributed and centralized units (DUs/CUs) in the access network.\footnote{For the purpose of simplification, Fig.~\ref{fig:Architecture2}, shows the RAN, DU, and CU as a single entity.}

A true Internet approach, however, where the end points could belong to any operator could yield large latencies due to the vast transport network fiber infrastructure. Indeed, the desired signals need to travel through one operator's transport backhaul network and then reverse through the other operator's network. In practice, multiple operator networks are connected to each other at nominated points of interconnect. Therefore, a call (data or otherwise) from an end point of operator A to an end point of operator B must traverse through these nominated points of interconnects. Assuming a typical transport network backhaul of 300-500 km, this adds to between 600-1000 km fiber. Given the finite and reduced speed of light in fiber, this mounts to approximately 3-6 ms added latency in a non-congested transport network. Furthermore, in a congested network without a sliced architecture, an even larger delay occurs. Therefore, 5G in multi-operator environments is unable to offer the anticipated ultra-low latency QOS assurances for flexible network deployments, where the end points can belong to different operators. 

In order to address this, we propose the idea of a ``local breakout network", as shown in the lower half of Fig.~\ref{fig:Architecture2} labeled as ``data network". This will facilitate the reduction or removal of the transport network fiber infrastructure, and thereby reduce latency. However, it is also required for the general internet service provider (ISP) infrastructure to be ``slicable". An end-to-end orchestration approach is thus needed in 6G which would enable such deployment scenario. This orchestrator could be implemented on a distributed ledger to increase transparency between competing parties, as in Fig.~\ref{fig:Architecture2}.
\begin{figure*}[!t]
	\centering
	\hspace{25pt}
	\includegraphics[width=14cm]{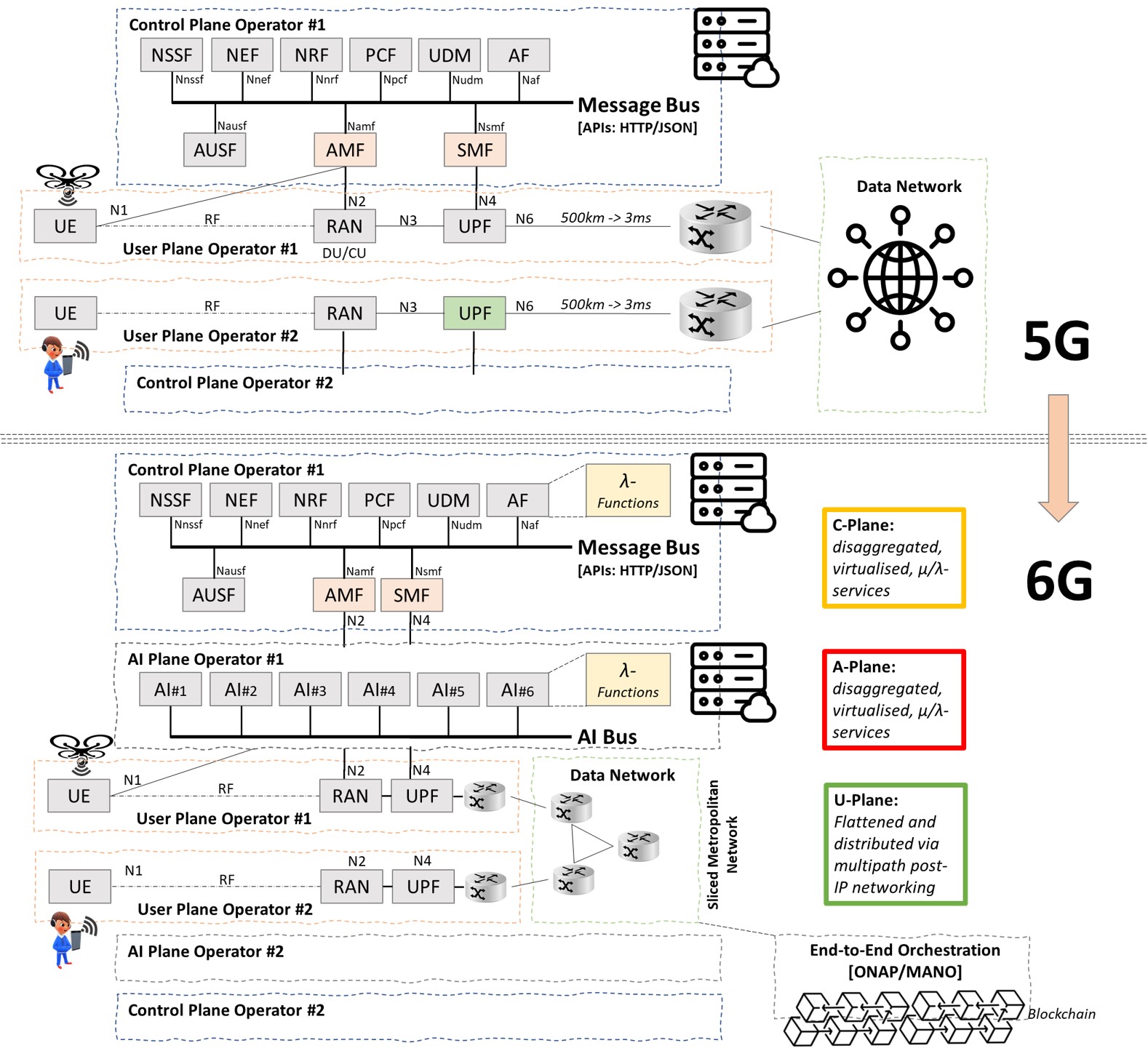}
	\vspace{-5pt}
	\caption{High-level overview of the 6G architecture, where: i) Compute/storage/networking has been flattened, ii) The transport network has been ``shortcut" with a sliced local breakout to enable low latency between the networks of two operators, iii) An AI-Plane (A-Plane) has been introduced in addition to a User-Plane (U-Plane) and Control-Plane (C-Plane). Furthermore, the 3GPP logical network entities such as PCF/AMF/UPF are being disaggregated further through cloud-centric lambda functions.}
	\label{fig:Architecture2}
	\vspace{-15pt}
\end{figure*}
\begin{table*}[!t]
\centering
\vspace{5pt}
\scalebox{0.66}{
\begin{tabular}{ccc}
\toprule 
\textbf{} & \textbf{Design Challenge} & \textbf{Opportunity} 
\tabularnewline
\midrule
\midrule 
\textbf{Super-Convergence}&\hspace{-48pt}\tabitem Technology interoperability& \hspace{-71pt}\tabitem Flexible traffic onboarding/offloading\\ &\hspace{20pt}\tabitem Flexible security/authentication mechanisms & \hspace{-56pt}\tabitem Higher reliability due to link redundancy\\
& & \tabitem Lower latency due to minimization of re-transmissions\\ \midrule
\textbf{Non-IP Based Networking}& \hspace{-62pt}\tabitem Sufficiently flexible and &\hspace{-33pt}\tabitem Networking adapted to modern network traffic\\ &\hspace{-30pt}generic networking protocol& \hspace{-23pt}\tabitem Facilitation of ICNs and citizen-centric networks\\
&\hspace{45pt}\tabitem Wider adoption of any viable protocol candidates & \\ \midrule
\textbf{Information/Intent-Based Networking} & \tabitem Insufficient, incomplete or sparse data & \hspace{-50pt}\tabitem Translation of intents to actionable design\\
& \hspace{10pt}\tabitem Too much redundant or our of time data & \hspace{-50pt}and/or parameterization of the network\\ \midrule
\textbf{360 Cyber Security} & \hspace{10pt}\tabitem Constantly evolving architectures, hence &\hspace{-38pt}\tabitem Development of all encompassing framework\\ 
& \hspace{13pt}constantly evolving surfaces of attack&\hspace{-55pt}including design and implementation\\ 
&\hspace{3pt}\tabitem{Unliked communities e.g., telco design} & \hspace{-85pt}\tabitem{Novel dynamic security principles}\\
&\hspace{-25pt}and embedded programming &\\ \midrule
\textbf{Privacy by Engineering Design} & \hspace{-6pt}\tabitem Translate preferences and contextual&\hspace{-23pt}\tabitem Bake privacy natively into the infrastructure, thus \\ 
&\hspace{2pt}information into a real-time design & \hspace{7pt}circumventing problems with privacy by T\&C design\\ \midrule 
\textbf{Future-Proving Emerging Technologies} & \hspace{10pt}\tabitem Difficult to gauge all future technologies & \hspace{-40pt}\tabitem Enable a more dynamic and atomized design \\ 
& \hspace{49pt}\tabitem Danger of patch work from a technical viewpoint& \hspace{-40pt}going well beyond microservice thinking\\
\midrule 
\textbf{Transport Network Removal/Reduction} &\hspace{-10pt}\tabitem Legacy system and legacy thinking & \hspace{-67pt}\tabitem Re-focus the wireless community and \\ 
& \hspace{-5pt}\tabitem Not all technology needed to enable & \hspace{-40pt}industry on the main focus: Wireless edge\\
& \hspace{70pt}such a transformation is softwarized and upgraded & \hspace{-65pt}\tabitem Much higher efficiency and lower cost \\ \midrule 
\textbf{Flattened Compute-Storage-Transport} & \hspace{-10pt}\tabitem Legacy system and legacy thinking & \hspace{-83pt}\tabitem End-to-end system which is more  \\ 
& \hspace{-5pt}\tabitem Not all technology needed to enable & \hspace{-80pt}reliable and resilient to outages\\
& \hspace{70pt}such a transformation is softwarized and upgraded & \hspace{-35pt}\tabitem More cost-efficient and operationally effective \\
\midrule
\textbf{Native Open Source Support} & \hspace{-7pt}\tabitem Resistance from incumbant industry & \hspace{-70pt}\tabitem Significantly lower development costs\\ 
&\hspace{28pt}\tabitem Unstable or unsupported software principles& \hspace{-85pt}\tabitem Significantly lower time to market \\
&\hspace{-73pt}in the long-term& \hspace{-56pt}\tabitem Leverage on crowd skills and capabilities \\
\midrule
\textbf{AI-Native Design \& Human-Machine Teaming} & \hspace{25pt}\tabitem Algorithmic frameworks not mature enough& \hspace{-38pt}\tabitem Unique opportunity to advance meta-learning\\
& \hspace{-15pt}\tabitem Unstable and inoperable networks&\hspace{-65pt}for machine-driven system designs \\
& \hspace{-90pt}\tabitem Ethics and trust& \hspace{-28pt}\tabitem Create platform which is likely to be the future\\ 
& &\hspace{-86pt}of human-machine interaction\\ \midrule
\textbf{Human-Centric Networking}& \hspace{55pt}\tabitem Break the current network/device centric paradigm & \hspace{-45pt}\tabitem Establish a novel design paradigm which is\\
& \hspace{-55pt}\tabitem Privacy, ethics and trust & \hspace{-20pt}better suited in addressing societal challenges\\ \bottomrule
\end{tabular}}
\vspace{5pt}
\caption{A summary of the challenges and opportunities associated with disruptive designs of the 6G infrastructure.}
\label{6GDesignChallenges}
\vspace{-23pt}
\end{table*}

\vspace{-15pt}
\section{New Physical Layer Techniques for 6G}
\label{modulation}
We begin the section by discussing the current progress and future directions of modulation, waveforms and coding techniques essential for the next generation air interface design. This is followed by a detailed discussion on multiple antenna techniques spanning ultra massive MIMO systems, distributed antenna systems, intelligent surface-assisted communications and oribtal angular momentum (OAM)-based systems. We then discuss the state-of-the-art in multiple access techniques complementing the multiple antenna techniques. Motivated by THz frequencies, we analyze the realistic possibilities in free-space optical communications. Following this, we provide a discussion on the PHY applications requiring AI and ML. We conclude the section by discussing the current state of affairs and practical possibilities in vehicular communications. For space reasons, we do not present other important topics such as dynamic spectrum sharing, dual connectivity, full-duplex communication, as well as integrated access and backhaul. Readers can refer to \cite{MOSLEH1,ERICSSONTECHR1,ERICSSONTECHR2} for a discussion on these topics.

\vspace{-10pt}
\subsection{Modulation, Waveforms, and Codes}
\subsubsection{\underline{Multicarrier Techniques}}
Over the past decade, orthogonal frequency-division multiplexing (OFDM) has, by-far, become the most dominant modulation format. It is being applied in the downlink for both 4G and 5G, while the uplink could either be discrete Fourier transform (DFT)-precoded OFDM (for 4G and optionally for 5G) or conventional OFDM (5G). OFDM's popularity is rooted in two factors: 1) Its well-known \emph{information-theoretic optimality} for the maximization of system capacity over frequency selective channels. 2) \emph{Backward compatibility} - OFDM was chosen as a modulation method for 4G, and as a result has also been employed in 5G. While the trend in 5G has been the unification of modulation formats to OFDM for its three major use cases adapting the \emph{numerology} and \emph{frame structure}, we anticipate that the increased heterogeneity of applications in 6G will bring a much wider range of modulation formats. In particular, those which are suitable for the various edge cases of 6G systems, such as massive access from IOT devices and Tbps directional links. Having said this, for some 6G applications, OFDM may still be \emph{retained} due to backward compatibility. Nonetheless, it has long been pointed out that OFDM has a number of drawbacks arising in non-ideal situations, which motivates further research into either modified multicarrier systems, or other alternatives.\footnote{In fact, a number of such other techniques were already explored for 5G, but their adoption was hindered by the tight standardization schedule for making 5G commercially operational.} 

The three key challenges of OFDM are: 1) Sensitivity to \emph{frequency dispersion}, 2) Reduction of spectral efficiency due to the cyclic prefix that combats \emph{delay dispersion} effects, 3) High peak-to-average power ratio (PAPR). All of these effects are becoming more critical at mmWave and THz frequencies, since frequency dispersion increases due to the higher Doppler shifts and phase noise. However, combating its effects by increasing the subcarrier spacing would reduce spectral efficiency due to the cyclic prefix (contrary to the popular opinion: delay spreads do not \emph{decrease} significantly with carrier frequency, though the small cell sizes and strong beamforming typically used at high frequencies might reduce it). In particular, \emph{interference} between the subcarriers of different UEs inevitably reduces performance of OFDM. High PAPR drives the requirement for highly linear power amplifiers (PAs) and high resolution data converters, e.g., analog-to-digital (ADC) and digital-to-analog (DAC) converters. This proves to be highly problematic since PAs need to operate with high backoff powers sacrificing their efficiency; and the energy consumption of ADCs/DACs becomes too high. The ADC/DAC resolution scales with bandwidth, making their design increasingly difficult and expensive. To this end, investigation into modulation techniques which strike the right balance between optimality of capacity and ADC/DAC resolution are required, keeping in mind the maximum admissible complexity in the equalization process \cite{landau2018achievable}. The equalization methods could also include re-configurable analog structures. In this line, a promising method is given by the temporally oversampled zero-crossing modulation, where information is encoded in the \emph{temporal distance between two zero crossings} \cite{fettweis2019zero}. As shown in Tab.~\ref{modulationformats}, a number of other modulation methods have been introduced, which can be classified into \emph{orthogonal}, \emph{bi-orthogonal} and \emph{non-orthogonal} categories. All of these methods fulfill any of the following three goals: 1) Enable a critically-sampled lattice, such that the symbols are centered in the time-frequency plane, leading to high spectral efficiency; 2) Achieve orthogonality in the complex domain to facilitate simple demodulation; and 3) Have pulses that are well-localized in the time-frequency plane. As shown by the Balian-Low theorem in Fourier analysis, the three conditions cannot be fulfilled at the \emph{same} time. Tab.~\ref{modulationformats} summarizes the various trade-offs and offers means of comparison to analyze the capability of each method. Besides classical OFDM, other orthogonal techniques include null suffix OFDM, filtered multitone (FMT), universal filtered multicarrier (UFMC), lattice OFDM and staggered multitone (FBMC), see, e.g.,  \cite{9249568} and references therein. Among bi-orthogonal methods, there exists cyclic prefix OFDM, windowed OFDM, and Bi-orthogonal 
frequency-division multiplexing (FDM). For non-orthogonal schemes which need to eliminate inter-symbol interference via more complex receivers include generalized FDM (GFDM) and faster-than-Nyquist signalling. 

In contrast to the above, an alternative method that is recently developed is known as orthogonal time frequency space (OTFS) \cite{hadani2017orthogonal}. OTFS performs quadrature amplitude modulation (QAM) not in the time-frequency domain, but rather in the \emph{delay-Doppler} domain. This allows us to \emph{exploit} frequency dispersion as a source of diversity. Furthermore, OTFS allows for a much more flexible and efficient multiplexing of UEs with different power delay profiles and Doppler spectra. While real-time prototypes for OTFS already exist, further investigations of efficient equalization architectures, in particular for multiple antenna systems, as well as other real-time implementation aspects constitute an important research topic for the future. Yet another alternative to the above techniques is the use of \emph{non-coherent} or \emph{differentially coherent} detection. While non-coherent multiple antenna systems have been explored since early 2000s following the seminal work of Hochwald and Marzetta \cite{hochwald2000unitary}, recent efforts have been devoted to develop suitable detection methods for multiple antenna systems with large antenna arrays. Nonetheless, further research into optimization of the trade-off between complexity and performance is required. Since 6G is expected to provide a unified framework of even more diverse applications than 5G, modulation techniques that require extremely low energy consumption, such as for massive connectivity via IOT or internet-of-bio-things deserve further attention. Often in such applications, a remote ``node" operates on \emph{energy harvesting}, or must survive for years in a single battery charge. While  theoretical investigations have shown that \emph{flash} signalling to be optimal, it is practically infeasible, since it requires high PAPR and precise synchronization. For such applications, new modulation methods that minimize the \emph{total} energy consumption for the transmitter (for the uplink) and receiver (for the downlink) are required.\footnote{This includes the possibility that \emph{different} modulation methods can be used for uplink and downlink.} Research in this line will include \emph{clock-free receivers}, since the clock and clock distribution can constitute a significant ``floor" in the overall energy consumption. To this end, even a ``sleep mode" requires significant energy if the clock needs to run to determine when the device needs to ``wake-up". Here related ideas from molecular communications may prove to be useful \cite{li2019clock}, as well as the use of passive backscatter communication \cite{van2018ambient} which help in improving the energy efficiency of devices. 

A major challenge of all modulation formats is the acquisition of channel state information (CSI) at the receiver and at the transmitter. In 5G systems, pilot signals are transmitted, which are later used for channel estimation at the receiver. For CSI acquisition at the transmitter, the system either relies on  \emph{reciprocity}, or \emph{feedback} from the receiver \cite{6375940}. However, the pilot overhead can become prohibitive, particularly in systems employing a massive number of transmit and/or receive antennas. Hence, for 6G, better CSI acquisition schemes need to be determined. Promising research efforts in this area include adaptation of the pilot signal spacing in time and frequency domain, exploitation of limited angular spread of the channel \cite{adhikary2013joint}, and advanced signal processing methods for reduction of pilot contamination \cite{muller2014blind}. A related problem is the quantization and feedback of CSI. While reciprocity calibration, and spatial/temporal extrapolation have shown promise  \cite{rogalin2014scalable}, system-level imperfections limit the desired gains, where further research needs to be conducted. Here, ML applications are worthy of exploration \cite{hoydis2020}.
\begin{table}[!t]
\centering
\scalebox{0.64}{
\begin{tabular}
[c]{llll}\toprule
\hspace{25pt}\textbf{Modulation} & \hspace{15pt}\textbf{Complex} & \hspace{-5pt}\textbf{Critically Sampled} & \textbf{Well-Localized}\\ 
\hspace{27pt}\textbf{Technique}& \hspace{5pt}\textbf{Orthogonality} & \textbf{Sampled Lattice} & \hspace{-5pt}\textbf{Localized Filters}\\\toprule\midrule
\hspace{35pt}\textbf{OFDM} & \hspace{30pt}Yes & \hspace{29pt}Yes & \hspace{30pt}No\\\hline
\hspace{30pt}\textbf{NS-OFDM} & \hspace{30pt}Yes & \hspace{30pt}No & \hspace{30pt}No\\\hline
\hspace{30pt}\textbf{FMT \cite{cherubini2002filtered}} & \hspace{10pt} Depends \cite{mattera2020windowed} & \hspace{26pt} No & \hspace{25pt} Yes\\\hline
\hspace{25pt}\textbf{Lattice OFDM} & \hspace{32pt}No & \hspace{29pt}No & \hspace{28pt}Yes\\\hline
\hspace{15pt}\textbf{Staggered Multi-tone} & \hspace{32pt}No & \hspace{29pt}Yes & \hspace{28pt}Yes\\\hline
\hspace{30pt}\textbf{CP-OFDM} & \hspace{30pt}Yes & \hspace{30pt}No & \hspace{30pt}No\\\hline
\hspace{15pt}\textbf{Windowed OFDM} & \hspace{15pt}Depends  \cite{mattera2020windowed} & \hspace{30pt}No & \hspace{15pt}Depends \cite{mattera2020windowed} \\\hline
\hspace{10pt}\textbf{Bi-Orthogonal OFDM} & \hspace{30pt}Yes & \hspace{30pt}No & \hspace{30pt}Yes\\\hline
\hspace{40pt}\textbf{GFDM} & \hspace{30pt}No & \hspace{30pt}No & \hspace{30pt}Yes\\\hline
\end{tabular}}
\vspace{13pt}
\caption{A qualitative comparison of contending modulation formats.}
\vspace{-23pt}
\label{modulationformats}
\end{table}

\subsubsection{\underline{Advances in Coding}}
In addition to novel modulation and waveforms, new codes also need to be designed. This is particularly the case for applications which require \emph{short packets}, such as in IOT systems. Low Density Parity Check (LDPC) codes and Polar codes that have \emph{short block lengths} have been employed for 5G systems for use in traffic and control uplink/downlink channels, respectively. The information-theoretic basis of achievable packet error rate as a function of block length has been established in \cite{polyanski2010}. On the one hand, codes with short block lengths are less reliable, such that error-free transmission cannot be easily guaranteed \cite{liva2016code}. An increase in the error probability may increase the need for automatic repeat request (ARQ) re-transmissions, which may not be suitable for time sensitive applications requiring ultra low-latencies. On the other hand, codes with longer block lengths also imply increasing latency. To this end, the interplay between the minimum required block length and robustness against transmission errors needs to be optimized keeping in mind the 6G KPIs listed in Tab.~\ref{table1}. Furthermore, low energy applications are often not well suited to ARQ, since this requires leaving the device in a non-sleep mode for an extended period of time, leading to an increase in energy consumption. New coding strategies should encompass both forward error correction and include novel iterative re-transmission/feedback mechanisms \cite{lentmaier2} and ML-based methods \cite{gruber2017deep}.\footnote{It is noteworthy that for long codewords in channels without fading, existing methods such as turbo decoding and belief propagation are highly efficient, operating close to the theoretical limits \cite{661103}.}

\vspace{-10pt}
\subsection{Multiple Antenna Techniques}
\label{MultipleAntennaTechniques}
\subsubsection{\underline{Ultra Massive MIMO Systems}}
The use of large antenna arrays have been one of the defining features of 5G systems. We foresee this trend to continue towards 6G systems, where the number of antenna elements will be scaled up by a further order-of-magnitude. The fundamental advantages of large antenna arrays have been discussed in overview papers for the past seven years \cite{heath2018foundations,marzetta2016fundamentals,SHAFI1,shafi20175g,bjornson2017massive}. A number of new research topics have emerged for study, which could prove valuable for 6G research. Firstly, the question of optimal beamforming architecture arises. For 5G deployments within the C-band, i.e., around 3.4-3.8 GHz, \emph{digital beamforming} remains a popular choice \cite{SHAFI2}, due to its ability to provide a higher beamforming gain, while utilizing the channel's spatial degrees-of-freedom \cite{marzetta2016fundamentals}. In sharp contrast, most current commercial deployments at mmWave frequencies, i.e., around 24.5-29.5 GHz, use \emph{analog beamforming} to explicitly steer the array gain in desired directions \cite{SHAFI2}. This is since digital beamforming at mmWave  frequencies yields high circuit complexity, energy consumption and cost of operation. Having said this, recent progress in high frequency electronics has facilitated digital beamforming for 64 antennas at 28 GHz as shown in \cite{DBF2}. In the future, closer investigations of fully digital implementations at mmWave frequencies are merited \cite{puglielli2015design}. In addition, the compromise solution of \emph{hybrid beamforming}, first proposed in \cite{zhang2005variable}, strikes the right balance between processing in the analog and digital domains has thus received considerable attention \cite{zhang2005variable,molisch2017hybrid,ABBASI1,TATARIAHYBRID1}. In Sec.~\ref{RFTransceiverChallengesandPossibilities}, we provide a more detailed discussion on the real-time processing and transceiver design trade-offs. Secondly, the impact of electrically ultra massive arrays arises as an important research direction. Most current massive MIMO implementations have limited electrical dimensions: for instance, a 256 element array might extend at most 8 wavelengths in one direction. However, as the dimensions increase even further, effects such as \emph{wavefront curvature} due to scattering in the near-field of the array, \emph{shadowing differences} in different parts of the array \cite{gao2015massive}, and beam squinting due to the non-negligible run time of the signal across the array \cite{wang2018spatial}, start to become much more pronounced. All of these physical artifacts need to be taken into account in the design and implementation of beamforming architectures and signal processing algorithms at the transmitter and receiver. Algorithms which provide the right balance between run-time complexity, ease of real-time implementation and optimality in performance need to be investigated, such as \emph{spatial modulation} - a lower complexity alternative to traditional multiple antenna methods. Here the \emph{index} of antennas are used to communicate part of the coded symbol \cite{di2013spatial}. More recently, various aspects of such methods are investigated for channel estimation, differential implementation and hybrid methods \cite{ishikawa201850}. Spatial modulation has also drawn interest at very high frequencies, such as those used for visible light communication (VLC) \cite{ZENG1} (see also Sec.~\ref{FreeSpaceOpticalCommunications}). An example of an ultra massive MIMO array is shown in Fig.~\ref{UltraMassiveMIMOFigure}. Either a single array from the 4096 elements can be formed, or multiple sub-panels (arrays) can be configured as shown in the figure. Naturally, the total number of elements is a design parameter and is subject to link budget considerations.

Progress in \emph{distributed antenna systems} has also been tremendous during the past five years, see e.g., \cite{NGOD1,BJORNSSOND1,SIG-109} and references therein. The concept of \emph{cell-free massive MIMO} has been pointed out as a promising way to realize distributed antenna systems below 6 GHz which can \emph{scale} to large physical areas. While the spectral and energy efficiency improvements brought by such systems are now well understood in theory, it remains to be seen whether the promised theoretical gains can be retained in practice for realistic scenarios with distances spanning up to hundreds of meters and variations in UE/scatterer mobility (see also Sec.~\ref{PropagationChannelsforDistributedAntennaSystems}). Since UEs can communicate with multiple access points at the \emph{same} time, a major research challenge in real-time implementations is to maintain synchronization between many distributed access points, UEs and the central processing unit.
\begin{figure}[!t]
    \centering
    \vspace{-10pt}
    \hspace{-32pt}
    \includegraphics[width=9.95cm]{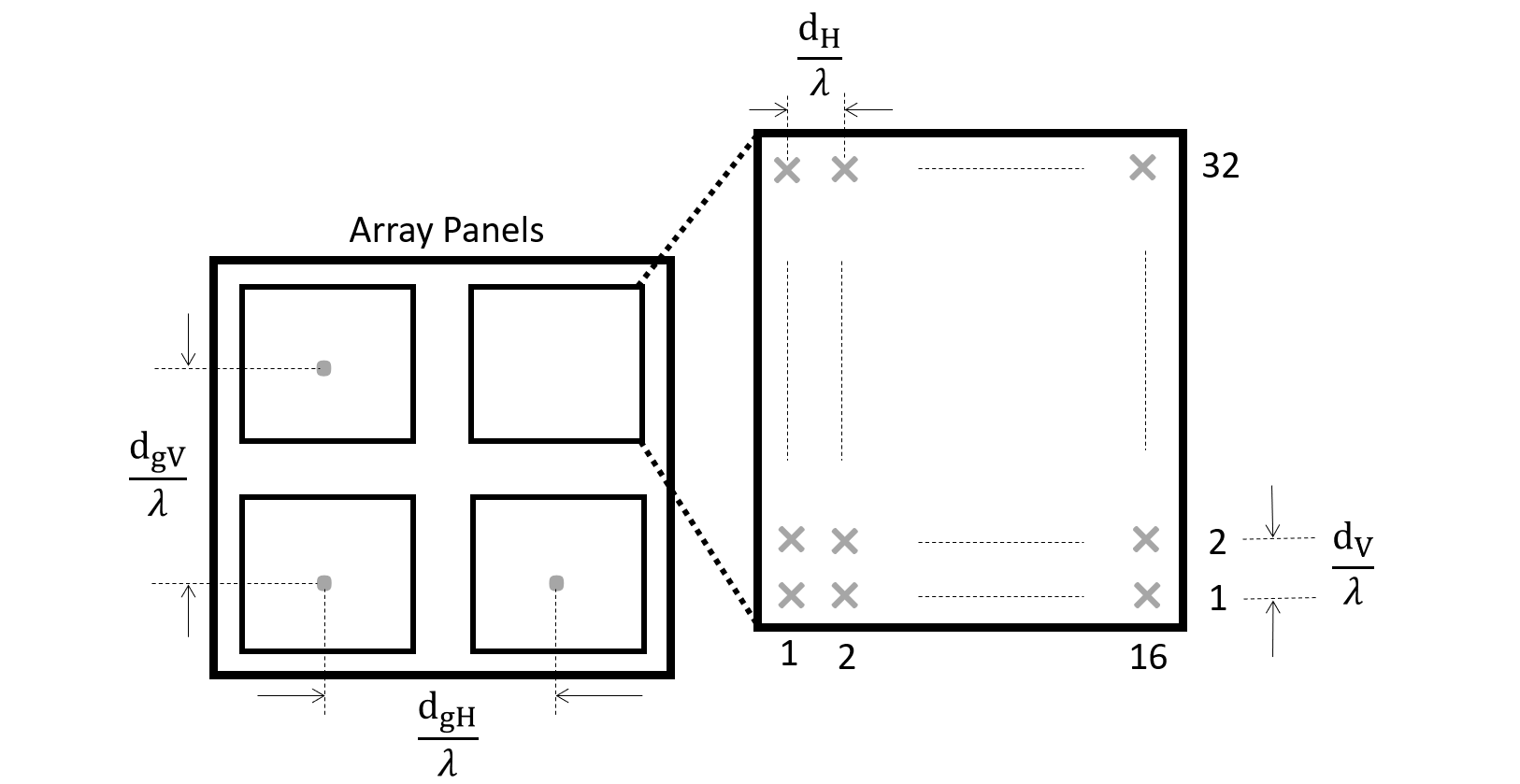}
    \vspace{-19pt}
    \caption{An example of an ultra massive MIMO array, consisting of four sub-panels, where each sub-panel has 32$\times$16 cross-polarized elements. Across all four panels, there are a total number of 4096 individual elements. The cross-polarized antenna elements in a sub-panel are spaced by $d_{\textrm{H}}/\lambda$ and  $d_{\textrm{V}}/\lambda$, where $\lambda$ is the operating wavelength. On the other hand, the panels are spaced by $d_{g\textrm{H}}/\lambda$ and $d_{g\textrm{V}}/\lambda$, respectively. Similar arrangements for conventional massive MIMO arrays are presented in \cite{3GPP1}.}
    \vspace{-1pt}
    \label{UltraMassiveMIMOFigure}
\end{figure}
\begin{figure}[!t]
    \centering
    \includegraphics[width=9.2cm]{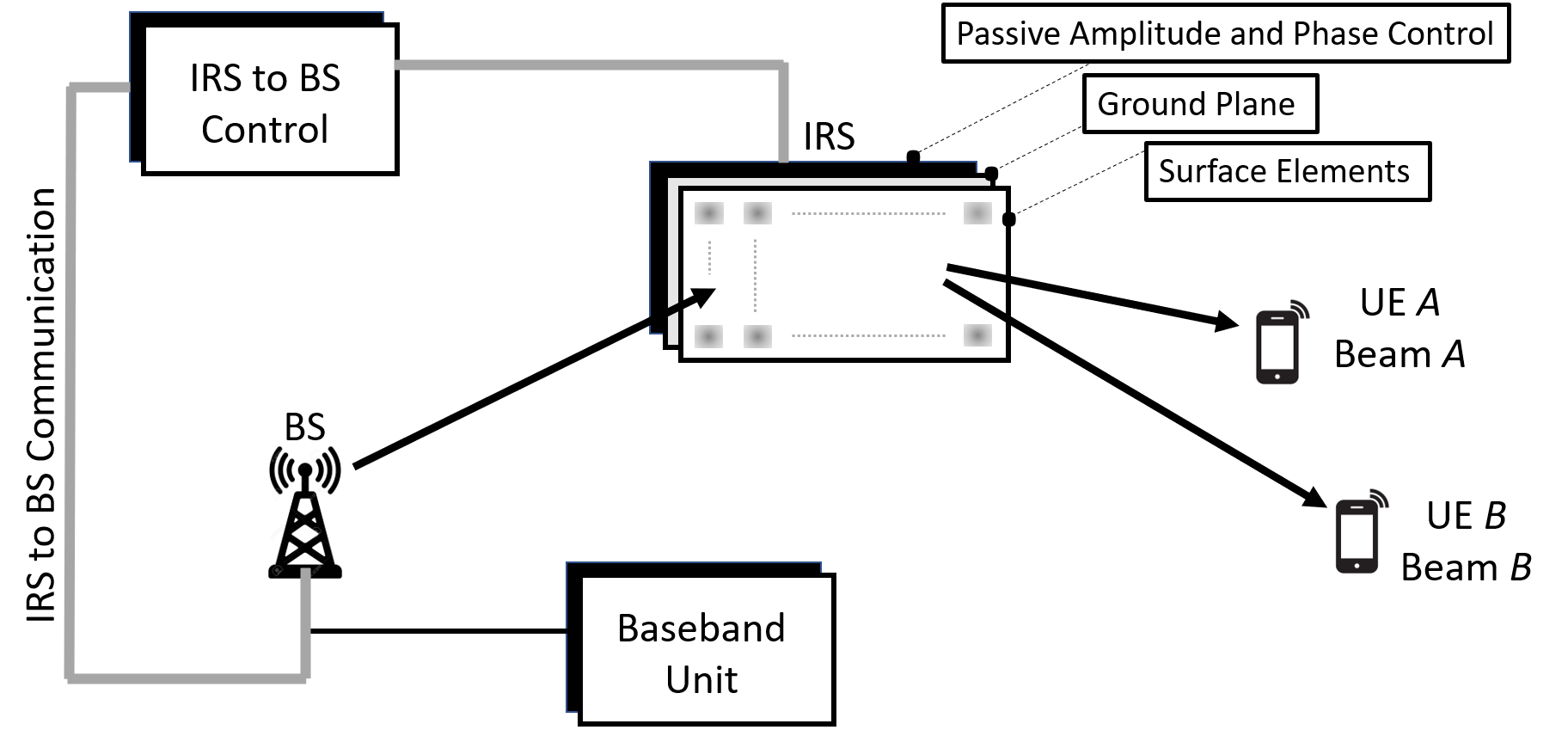}
    \caption{An intelligent surface-assisted communication system is shown, where a BS is communicating to two UEs (A and B) via an IRS. The IRS takes the shape of a planar array with elements distributed in the horizontal and vertical domains. The surface elements are coupled with a ground plane, followed by amplitude and phase control with passive electronic control networks. The IRS is intended to communicate with the BS via a dedicated controller.}
    \label{IRSOperation}
    \vspace{-14pt}
\end{figure} 
\subsubsection{\underline{Intelligent Surface-Assisted Communications}}
\label{IntelligentSurfacesAssistedCommunications}
Another important development is of \emph{large intelligent surfaces (LISs)} \cite{HU1,HU2}, which aim to have large physical apertures that are electromagnetically \emph{active}. The surface can be seen as an ultra massive MIMO array (as described above) capable of performing fully digital processing \cite{HU1,TATARIALIS1}. Browsing in the literature, many names of the same concept exist, such as \emph{re-configurable intelligent surfaces} and \emph{holographic beamforming} \cite{TATARIALIS1}. The inception of LISs led to the development of \emph{intelligent reflecting surfaces (IRSs)} \cite{TATARIALIS1,wu2019towards,9026400}, which are designed to \emph{quasi-passively} reflect the incoming signals to an adaptable set of outgoing directions via tunable phase shifters without any active down/up-conversion. A large number of papers are now appearing on both LISs and IRSs (see e.g., \cite{HU1,HU2,TATARIALIS1,zhao2019survey,yuan2020reconfigurable,elmeadawy2020enabling,wu2019towards} and references therein). In particular, for IRSs, a number of questions need further research, such as real-time steering and control of reflections, interference minimization and energy consumption optimization. Among the challenges whose solutions need to be researched are: 
\begin{itemize}
    \item Advantages and drawbacks relative to active relays, as well as to non-reconfigurable passive reflector structures. A \emph{joint} communications, electromagnetics and operation expenditure analysis needs to be carried out, which not not only considers the enhancement of coverage (which is frequency-dependent), but also circuit-level implications on performance, electromagnetic behavior of the array, and the actual cost of such deployments, such as the renting of space and ongoing maintenance. As such, guidelines for efficient deployment can be developed. 
    \item A detailed assessment on the \emph{reliability} of such structures needs to be carried out, which includes an analysis of the impact of possible ``pixel failures", i.e., elements in the array that do not operate due to cracks and/or other environmental factors, such as large variations in temperature, rain and wind. Potential solutions also need to take into account the impact of \emph{beam misalignment} due to these factors. Consequently, from an operator's viewpoint, maintenance of the surface elements and associated circuitry could be a significant overhead. 
    \item A detailed investigation into the control protocols to implement efficient signalling between the BS and the surface, as well as UEs and the surface need to be studied. In this line, important questions need to be answered, such as: how will the surface response be maintained with massive changes in radio traffic conditions due to e.g., handovers? How will insertion of the surface influence the design of core networks? Furthermore, novel algorithms for re-calibration on-the-fly need to be developed, or the IRSs need to be designed a-priori to work without any calibration, i.e., purely based on online pilot tones. 
    \item The backplane complexity of the surface relative to its aperture also needs to be investigated in detail.
\end{itemize}}
{\color{black}An example of an IRS-assisted communication system is shown in Fig.~\ref{IRSOperation}. The size of the IRS, i.e., the number of elements, along with its configuration and radiation efficiency are naturally its design parameters. A big challenge here is the dedicated link between the BS to the IRS; which  presents itself as an additional transport network of unspecified bandwidth, as this would depend upon the number of elements on the IRS, the number of control bits per-element, and their refresh rates coupled with the transmission times of the data frame structure.  Naturally, the needed throughput would scale with the number of UEs served for multiuser operation, and the required number of surfaces per-cell is not clear. If a network architecture similar to 5G must be followed, the termination point of the transport network will also involve communication with the centralized unit, as this is typically where inter-site coordinated multipoint computations take place. As a consequence, this poses a significant design challenge. \emph{Note that the addition of these links will not contribute to the user plane latency, yet will contribute to an increase of control plane latency.}

\vspace{3pt}
\subsubsection{\underline{OAM-Based Systems}}
\label{OAMBasedSystems}
Besides conventional spatial multiplexing, which is the fabric of existing multiple antenna systems, \emph{OAM} \cite{yan2014high} is an alternative spatial multiplexing method that has shown great potential for 6G systems. This technique imposes ``twists" on the \emph{phases} of the propagating laser beams, such that modes with different amounts of twist are orthogonal to each other. They can be easily separated through analog means, such as spiral phase plates. OAM is especially suitable for line-of-sight (LOS) propagation, such as in data centers, and for wireless backhaul, and are limited in range since due their underlying principle for multiplexing only works in the radiating near-field of the antenna. Investigations on how to make such systems robust to practical impairments of multipath, misalignment of orientation, etc., are critical to increase their practical utility. While some preliminary work has been done in that direction, e.g., in multipath propagation \cite{yan2016multipath} and turbulence \cite{lou2019new}, further work is required. Since OAM performs better with electrically larger antennas, it is better suited for high mmWaves and THz systems, and in particular for free-space optics applications, see e.g., \cite{willner2015optical,SAMSUNGWP1}.

\vspace{-11pt}
\subsection{Multiple Access Techniques}
\label{MultipleAccessTechniques}
\vspace{-1pt}
Multiple access techniques require a re-think in 6G, especially due to the integration of massive connectivity and extremely low energy applications. Current systems use carrier sense multiple access (CSMA) and/or non-contention access methods such as orthogonal time-frequency division multiple access for cellular systems. However, these multiple access schemes do not \emph{scale} well to scenarios where thousands of devices or more aim to access a single BS, but with a low duty cycle. Current work in this regime concentrates on spread-spectrum type approaches, such as long range (LoRa) communication, which results in low spectral efficiencies. Hence, new structures that allow for better scaling  and possibly further reduce latency need to be studied \cite{shirvanimoghaddam2017massive}. 

Another direction of future research is improvement of multiple access in the traditional high spectral efficiency approaches. Here, \emph{non-orthogonal multiple access (NOMA)} was originally intended to be part of 5G systems \cite{ding2017application}, yet was left out of the early releases due to the rush to finish the specifications. Another promising approach is known as \emph{rate splitting (RS)} \cite{CLERCKX1}. RS splits UE messages into common and private parts, and encodes the common parts into one or several common
streams, while encoding the private parts into separate streams. The streams are precoded using the available (perfect or imperfect) CSI at the transmitter, superposed and transmitted. All the receivers then decode the common stream(s), perform Successive Interference Cancellation and decode their private streams. Each receiver reconstructs its original message from the part of its message embedded in the common stream(s) and its intended private stream. The key benefit of RS relative to other techniques is to flexibly manage interference by allowing it to be partially decoded and partially treated as noise. We anticipate possible simplified versions of NOMA or RS to be in contention for 6G systems. In addition, 6G research should concentrate on how to further improve the performance up to the theoretical limits, while taking into account practical constraints on precoding, and amount of available CSI.

\vspace{-13pt}
\subsection{Free-Space Optical Communications}
\label{FreeSpaceOpticalCommunications}
More generally, free-space optical communications have great promise for extremely high data rate communications over small-to-medium distances, as long as LOS can be guaranteed. While some operations are possible also in  non LOS (NLOS) situations, the achievable data rates, and required modulation as well as signal processing structures can be quite different. To this end, more investigations are required to investigate architectures that provide the right complexity-cost-performance trade-off. We can generally distinguish between laser-based and light emitting diode (LED)-based techniques. The latter (a.k.a. VLC or LiFi) is mostly intended for exploiting LEDs that already exist as lighting source, for also transmitting information \cite{HAASx}. Furthermore, the optical transmission is intended for the downlink, while the uplink needs to be provided by traditional radio links. 
This raises interesting challenges in the integration with 6G cellular and 6G WiFi, which need much more attention. Furthermore, the adaptation to mobility constitutes an important challenge. Laser-based systems allow much higher data rates, yet having small beamwidths, they are mainly suitable for fixed wireless scenarios. Furthermore, they are extremely sensitive to blockage of the LOS paths, since no multipath diversity is available. Modulation and detection methods that are suitable in environments with fast variations of channel conditions also require further investigations.

\vspace{-12pt}
\subsection{Applications of AI and ML}
\label{ApplicationsofAIandML}
\vspace{-1pt}
A comprehensive survey of AI and ML applications for 5G and beyond is given in \cite{9023918}. For PHY research, ML techniques are currently being explored for a variety of tasks. Firstly, it can be used for symbol detection and/or decoding. While de-modulation/decoding in the presence of Gaussian noise or interference by classical means has been studied for many decades \cite{CAIRE1}, and optimal solutions are available in many cases, ML could be useful in scenarios where either the interference/noise situation does not conform to the assumptions of the optimal theory, or where the optimal solutions are too complex. {\color{black}Given the recent trend, 6G} will likely to utilize even shorter codewords than 5G (where Shannon theory does not hold) with low-resolution hardware (which inherently introduce non-linearity that is difficult to handle with classical methods). Here, ML could play a major role, from symbol detection, to precoding, to beam selection, and antenna selection. ML is generally very well suited for these PHY techniques, due to the large amount of training data that can be generated with comparatively little effort, and due to the ``labeled data" (ground truth) being readily available. Another promising area for ML is the estimation and prediction of propagation channels. Previous generations, including 5G, have mostly exploited CSI at the receiver, while CSI at the transmitter was mostly based on roughly quantized feedback of received signal quality and/or beam directions. In systems with even larger number of antenna elements, wider bandwidths, and higher degree of time variations, the performance loss of these techniques is non-negligible. Here ML may be a promising approach to overcome such limitations, see e.g., \cite{channelest}. {\color{black}In particular, questions related to the best ML algorithms given certain conditions, required amount of training data, transferability of parameters to different environments, and improvement of explainability will be the major topics of research in the foreseeable future.}

\vspace{-12pt}
\subsection{Vehicular Communications}
\label{VehicularCommunications}
\vspace{-2pt}
Modern vehicles are equipped with up to 200 sensors, requiring much higher data rates \cite{7786130}. Vehicles may also be equipped with video cameras, infrared cameras, automotive radars, light detection and ranging systems, as well as global positioning systems. The sensors and additional devices provides an opportunity to collaborate and share information in order to facilitate accurate and safer automated driving; particularly in congested scenarios. The raw aggregate data rate from the above sensors could be up to 1 Gbps, which is well beyond the capability of digital short range communication (DSRC) - the current protocol for connected vehicles \cite{5GAA}. Moving forward, we see the utilization of bands below 6~GHz for high reliability and mmWave bands to achieve Gbps data rates \cite{7742901,8338071}. {\color{black}Fundamentally, some important research challenges that need attention are: 1) Lack of accurate wave propagation models \cite{7742901,8338071}; 2) Assessment of the impact of cars through car penetration loss and antenna arrangements (see, e.g., \cite{Kato01,He20}); 
3) Lack of accurate modeling of channel non-stationarities (see, e.g., \cite{Abbas13}). 
On the network side, we predict that the current 5G network architecture will not meet the latency needs of reliable autonomous driving until MEC is fully integrated. Besides the channel and network aspects, for V2X scenarios, a large number of PHY-related questions need to be investigated. In particular, the processing of the sensor data, including \emph{sensor fusion}, will become a major bottleneck due to the combination of large amount of data and tight processing deadlines. The optimal trade-offs between processing at the point of origin, at the BS (if involved), and at the end-point need to be determined, taking into account its relationship with a given level of traffic density, the amount of available infrastructure, as well as the real-time computational capabilities of involved cars. We expect that much of information fusion will occur via ML algorithms. With high mobility of cars and blockages by intervening vehicles, \emph{beam management} is another aspect which needs much more research. In particular, the \emph{beam adjustment} mechanisms designed for 5G are often too slow in adapting to vehicular scenarios, calling for new methods. For V2X/V2I systems, the fast association/disassociation with the various road-side units may require a distributed antenna deployment (discussed further in Sec.~\ref{PropagationChannelsforDistributedAntennaSystems} from a propagation aspect), and its implications on PHY need to be studied. 

Importantly, even if all of these research challenges can be addressed, it must be noted that the large number of old cars on the roads will limit the true gains of V2X/V2I systems until late 2030's when the majority of the cars may have V2X/V2I capability.} Combinations of DSRC, long-term evolution (LTE), cellular V2X and mmWaves offers a unique opportunity to simultaneously improve reliability, data rates and intelligence of vehicular networks \cite{3GPP1}.
{\color{black}Figure \ref{fig:V2X} shows an example of a 6G vehicular use case. High rate low-latency mmWave links are deployed within the platoon at the bottom of the figure for cooperative sensing and sensor fusion. They are also deployed between the vehicles and a controller connected to a uRLLC network is shown on the bottom right for sensor sharing and vehicle control. A sub-6 GHz network is used by all vehicles to broadcast basic information such as Cooperative Awareness Messages (including the position and velocity vector), and for intersection control in busy scenarios to improve efficiency. Vulnerable road users, being equipped with communication devices or not, are protected through collective perception based on sensor data from the infrastructure (e.g. the cameras in the middle of the figure) as well as from vehicles.
}
\begin{figure}[!t]
    \centering
    \hspace{40pt}
    \includegraphics[width=8.5cm]{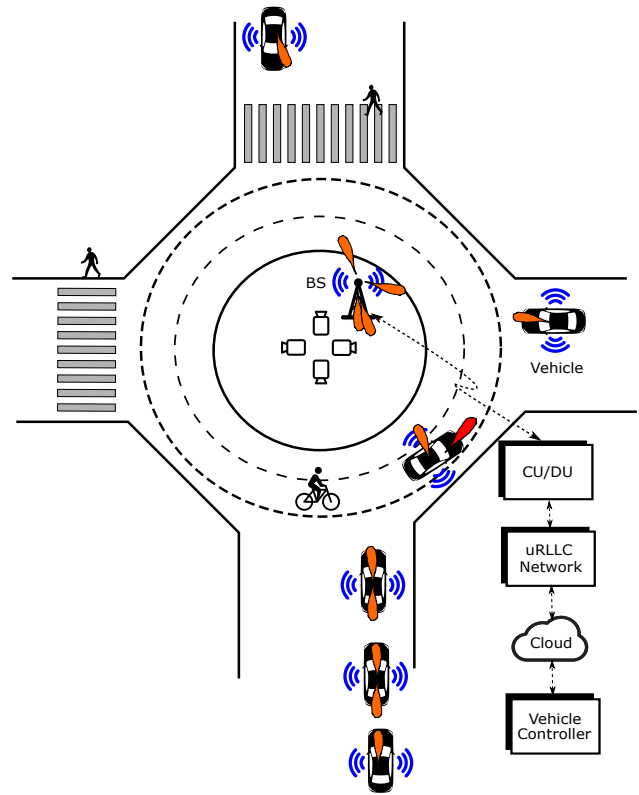}
     \caption{{\color{black}Vehicular use case: Here a roundabout with high rate low latency mmWave links for sensor sharing with a vehicular controller through a uRLLC network (from the middle of the figure to the bottom right), and within the platoon for cooperative sensing (bottom). Sub-6 GHz communication is used for basic information such as Cooperative Awareness Messages. The BS shown has multiband capability, and its interface to the core network is via the distributed and centralized units (DU/CU), see the bottom right.}}
    \label{fig:V2X}
   \vspace{-15pt}
\end{figure} 
Since 6G services are expected to be planned over an extremely wide range of frequencies, we now review the propagation characteristics over which 6G systems will operate. 

\vspace{-6pt}
\section{Propagation Characteristics of 6G Systems} 
\label{models}
The performance of 6G systems will ultimately be limited by the propagation channels they will operate over. It is thus of vital importance to investigate the propagation characteristics relevant for 6G systems, in particular those that have not already been explored for earlier-generation systems. This section provides an overview of wave propagation mechanisms for sub-6 GHz, mmWave and THz frequencies. Across these frequencies, we characterize ultra massive MIMO channels, distributed antenna channels, V2V and V2I channels, industrial channels, UAV channels, and wearable channels, respectively. Other important topics such as full-duplex channels and device-to-device channels are omitted due to space reasons. Interested readers can refer to \cite{shafi20175g,SHAFI1}, as well as references therein, for a more comprehensive overview.

\vspace{-10pt}
\subsection{MmWave and THz Propagation Channels}
\label{mmWaveandTHzPropagationChannels}
Moving to new frequency bands usually entails determination of the fundamental \emph{propagation processes}. As has long been pointed out by wireless textbooks, using constant gain antennas, the free-space pathloss {\em increases} with $f^2$, where $f$ is the carrier frequency, and {\em decreases} with $f^2$ when \emph{constant-area} antennas are used at both link ends. As such, for a given form-factor, highly directional antennas can provide low free-space pathloss. This has driven the need for massive MIMO arrays at mmWave frequencies and ultra massive MIMO arrays THz frequencies. In the mmWave bands, the atmosphere can become absorbing (depending on $f$), attenuating the received signal as $\textrm{exp}(\alpha_{\textrm{atm}}d)$, where $d$ is the distance between the BS and UE. The attenuation coefficient, $\alpha_{\textrm{atm}}$, is a function of $f$, as well as the atmospheric conditions, such as fog, rain, etc. \cite{SMITH1}. As depicted from Fig.~\ref{AbspPeaks}, atmospheric attenuation in the THz bands is much higher than the mmWave bands. Notably, the only strong attenuation below 100 GHz is the oxygen line at 60 GHz, giving rise to a loss of approximately 10 dB/km; while from 100-1000 GHz, multiple attenuation peaks exist, that can exceed 100 dB/km. 
The physical origin of this absorption - a.k.a. molecular absorption - is that electromagnetic waves of specific frequencies excite air molecules causing internal vibration, during which part of the energy driving the propagating wave is converted to kinetic energy and lost. 
{\color{black}\emph{The above discussions show that band selection must be carefully aligned with the anticipated distance between the BS and UEs.}}  

As seen by the Fresnel principle, the efficiency of \emph{diffraction} is greatly reduced at mmWave and even more at THz frequencies, since common objects introduce \emph{sharp} shadows \cite{SHAFI1}. On the other hand, \emph{diffuse scattering} becomes highly relevant, since the \emph{roughness} of surfaces (in terms of wavelengths) becomes considerable \cite{jansen2011diffuse}. Unlike the lower frequencies where it is common to assume that a plane wave incident on a rough surface results in a \emph{specularly reflected} wave and its diffuse components are scattered \emph{uniformly} into all directions, at THz bands, there is a general lack of validation of this concept via measurements. It is speculated that the amplitude of scattered paths may not be large enough to significantly contribute to the impulse response - an effect that is also observable at mmWave frequencies. Furthermore, attenuation by vegetation, as well as penetration losses in outdoor-to-indoor propagation increase dramatically at mmWave frequencies \cite{BASx}. Several studies have been conducted to better understand the material dependence on propagation characteristics for bands below 100 GHz, see e.g., \cite{MEDBO3,BAS1}. However, relatively fewer such studies exist for the THz bands, where a full assessment of reflection, transmission, and scattering coefficients of many building materials has been done only in few papers \cite{PRIEBE1}. Specular reflections at a \emph{dielectric half-space} (most commonly ground reflections) are frequency dependent so long as the dielectric constant is frequency dependent, while reflection at the \emph{dielectric layer}, such as a building wall, depends on the \emph{electrical thickness} of the wall, and thus on frequency. Having said this, it is not clear whether reflection coefficients increase or decrease with frequency. Conversely, power transmitted through objects  decreases almost \emph{uniformly} with frequency due to the presence of the skin effect in lossy media \cite{SHAFI1}. Last but not the least, Doppler shifts scale \emph{linearly} with frequency, while the first Fresnel zone \emph{decreases} with square root of the wavelength. 
For realistic simulations, all of these physical effects need to be incorporated into \emph{ray tracers} and \emph{statistical models}. {\color{black}Accurately accounting for the physical environmental features is a major challenge for ray tracing, as
well as obtaining sufficiently high-resolution databases of the terrain. It is known already that standard databases only offer resolution on the order of a few meters, and thus do not show effects such as critical transitions from smooth windows to rough stucco material, for example. Keeping in mind the above discussions, there exist several standardized and non-standardized models for impulse response generation in the mmWave bands, see e.g., \cite{HANEDA1,XING1,TATARIAX,SHAFI1,shafi20175g} for a detailed taxonomy and model parameters. However, the same cannot be said for THz bands due to largely unknown measured characteristics. There exist some recent investigations in  \cite{NABBASI1,PRIEBE1,xing2018propagation,rey2017channel} around 140 GHz and in the 275-325 GHz band, from which finite multipath component (MPC) models of the THz channels are derived. Notably, the authors in \cite{PRIEBE1} propose a relatively detailed hybrid model for indoor channels combining spatial, temporal and frequency domains with parameters from 140-150 GHz and 275-325 GHz, respectively. While most existing measurements are short-distance and/or for a fixed horn orientation, \cite{NABBASI1} provides the first outdoor, directionally resolved, measurements over longer (100 m) distance.}  

{\color{black}Going forward, a lot more work is required to remedy the lack of models for both high mmWaves and THz channels, with the main challenges being as follows: 1) Design and construction of suitable measurement equipment: even for mmWave channels, the construction of channel sounders with high directional resolution, large bandwidth, and high phase stability is very difficult, expensive and time consuming; the lack of available phased arrays and the low output power beyond 200 GHz make measurements even more difficult at those frequencies. Significant effort by the wave propagation community will be required to be able to perform large-scale measurements of static and dynamic channels. 2) Most current channel models are for very specific indoor scenarios, and the presence of a larger variety of environments, as well as different objects in the surroundings will require a \emph{mixed} deterministic-stochastic modelling approach \cite{PRIEBE1}. In order to characterize the stochastic part of the model, extensive measurements are required, which are currently missing, pointing to the large open gaps at THz frequencies.}

A summary of the key THz propagation characteristics and its impact on THz systems, as well as a comparison relative to lower bands is depicted in Tab.~\ref{THzDifferences}. 
\begin{table*}[!t]
\centering
\scalebox{0.71}{
 \begin{tabular}{c c c c} 
 \toprule
 \textbf{Parameter} & \hspace{-8pt}\textbf{Frequency Dependence} & \hspace{-18pt}\textbf{Impact on THz Systems} & \hspace{-38pt}\textbf{THz vs. Lower Bands} \\ [0.5ex]
 \hline\hline
 \textbf{Free-Space Pathloss} & \hspace{-18pt}\tabitem Increases with square of& \hspace{-30pt}Distances are limited to & \hspace{-1pt}Loss as a function of frequency\\
 & \hspace{19pt} $f$ when constant gain antennas & \hspace{-35pt}tens of meters at most &\hspace{-22pt}remains, hence THz loss is \\ 
 & \hspace{-68pt}are used & & \hspace{-40pt}higher than microwave \\ 
 & \hspace{-23pt}\tabitem Quadratic decrease with & & \\
 & \hspace{2pt}constant area \& frequency & & \\
 & \hspace{-44pt}dependent gain & & \\ \hline
 \textbf{Atmospheric Loss} & \hspace{-2pt}Absorption peaks that are & \hspace{-15pt}\tabitem Significant absorption loss. & \tabitem No clear effects at microwave\\
 & \hspace{1pt}dependent on frequency \& &  \tabitem Useful spectra limited between & \hspace{-16pt}\tabitem $O_2$ molecules at mmWave\\
 & \hspace{-50pt}$H_{2}\hspace{1pt}O$ contents & \hspace{-40pt}low loss windows& \tabitem $H_{2}\hspace{1pt}O$ \& $O_2$ molecules at THz\\ \hline
 \textbf{Diffuse Scattering \& Specular Reflections} & 
 \hspace{-12pt}\tabitem Diffuse scattering increases & \hspace{-35pt}Limited multipath \& 
& \hspace{-20pt}\tabitem Stronger than microwave \\ 
 & as a function of frequency & \hspace{-70pt} high sparsity  \\
 &\tabitem Frequency dependent specular & & \\
 & \hspace{-52pt}reflection loss & & \\ \hline
 \textbf{Diffraction, Shadowing and LOS Probability} & \hspace{-39pt}\tabitem Negligble diffraction & \hspace{-50pt}\tabitem Limited multipath & \hspace{-20pt}\tabitem Stronger than microwave \\ 
 & \tabitem Shadowing \& penetration loss & \hspace{-60pt} high sparsity & and mmWave frequencies\\
 & \hspace{-10pt}increases with frequency & \hspace{-40pt}\tabitem Dense spectral reuse & \\ \
 & \hspace{-10pt}\tabitem Frequency independent LOS & &  \\
 & \hspace{-70pt}probability & \\ \hline
 \textbf{Weather Influences} & \hspace{6pt} Frequency dependent airborne & Attenuation caused by rain & \hspace{-20pt}\tabitem Stronger than microwave \\
 & \hspace{-27pt} particulates scattering & & and mmWave frequencies\\ \bottomrule
 \end{tabular}}
 \vspace{15pt}
 \caption{THz wave propagation characteristics, impact on system performance and comparison with lower bands}
 \label{THzDifferences}
 \vspace{-20pt}
\end{table*}

\vspace{-9pt}
\subsection{Propagation Channels for Distributed Antenna Systems}
\label{PropagationChannelsforDistributedAntennaSystems}
6G systems will significantly evolve distributed BSs, in the form of either enhanced cloud RAN systems, coordinated multipoint transmission (CoMP, a.k.a. cooperative multipoint) or cell-free massive MIMO systems. As it currently stands, the majority of the deployments will be carried out for bands below 6 GHz. However, in order to complement the high reliability with high data rates, we foresee the use of mmWave bands, where not so many investigations exist.

For multiuser scenarios in either of the two bands, the \emph{joint} channel conditions for multiple UEs have need be provided. A greater challenge is the modelling links from a single UE to multiple BSs. Much of the earlier work has concentrated on the correlation of shadowing between different links. More recent measurement campaigns have quantified the correlation of parameters such as angular spreads, delay spreads, and mean directions \cite{ZHU1}. Typically, it is found that significant link correlation can exist even if the BSs are far away from each other; positive correlation can be found when the BSs are in the same direction from the UE. The correlation of BSs can be modeled through the concept \emph{common clusters}, i.e., clusters that interact with MPCs from different UEs as shown in \cite{POUTANEN1}. For instance, if these clusters are shadowed, it affects the net received power, as well as the angular and temporal dispersion of multiple UEs \emph{simultaneously}. This concept has been adopted in the design of the European Cooperation in Science and Technology (COST) 2100 channel model. 

\vspace{-11pt}
\subsection{Ultra Massive MIMO Propagation Channels}
\label{UltraMassiveMIMOPropagationChannels}
With the maturity of co-located and/or distributed massive MIMO systems, along with the emergence of LISs and IRSs, the number of radiating elements are foreseen to increase beyond those which are conventional today \cite{akyildizdistance,Jorneymodel,Akyildijorneyhan,HU1,HU2,TATARIALIS1}. Ultra massive MIMO arrays are primarily envisioned to operate at high mmWave and/or THz frequency bands, where potentially thousands of antenna elements can be integrated into small form factors \cite{akyildizdistance,Jorneymodel,Akyildijorneyhan}. The authors of \cite{akyildizdistance,Jorneymodel,Akyildijorneyhan} provide a taxonomy of ultra massive MIMO operation at THz frequencies using the \emph{arrays of subarrays} concept. 
Since antenna arrays at high mmWave and/or THz bands become \emph{physically small}, from a propagation viewpoint, they do not contribute to additional insights than those already described in the mmWave and THz propagation section, i.e.,  Sec.~\ref{mmWaveandTHzPropagationChannels}. 

In contrast, bands below 6 GHz also provide interesting research opportunities for ultra massive MIMO channels  \cite{HU1,HU2,BJORNSON1,DECARV1,ALI1}- though deployment of such large arrays at these frequencies is challenging. As the number of antenna elements are increased, the total \emph{physical} aperture of the radiating elements is also increased. As this happens, conventional propagation theories and results exploiting the plane wave assumption start to breakdown. Fundamentally, the \emph{Fraunhofer distance} denoted by $d_{\hspace{-1pt}f}$, is given by $d_{f}=2D^{\hspace{1pt}2}/\lambda$, where $D$ is the maximum dimension of the array and $\lambda$ denotes the wavelength. An increasing $D$ with a fixed $\lambda$ would imply that the UEs, as well as the scatterers would be increasingly likely to be within the \emph{Fresnel zone} of the antennas - one which corresponds to the radiating near field. This has some fundamental consequences on the overall propagation behavior. Firstly, \emph{spatial non-stationarities} in the channel impulse responses start to appear over the size of the array, where different parts of the array ``sees" (partially) unique set of scatterers and UEs \cite{GAO1,ALI1,TATARIA2,TATARIA3,tatariathesis,T1,T2}. As a consequence, the effects of wavefront curvature starts to vary not only the phases of the MPCs, but also the amplitudes \emph{over the array size}. To this end, the effectiveness of channel hardening and favorable propagation - two pillars of massive MIMO channels start to lose effect leading to increased variability in channel statistics \cite{LIICC,7996528}. Secondly, any propagation model to/from ultra massive MIMO arrays need to be directly linked to physics of near-field propagation to compute the near-field channel impulse response. A detailed procedure is given in \cite{HU1,HU2,BJORNSON1} to generate such response. 

Several measurement-based studies have demonstrated the above effects quantitatively, see e.g., \cite{GAO1,GAO2,DECARV1,ALI1}. The authors of \cite{GAO1,GAO2} show the effects of spatial non-stationarities from a 128 element virtual linear array (movement of a single element along the horizontal track) in outdoor environments at 2.6 GHz over a 50 MHz bandwidth. The array spanned 7.4 m with half wavelength spacing between the position of successive elements was serving a single UE in LOS or NLOS propagation. 
The authors in \cite{DECARV1,ALI1} report a similar measurement-based analysis of ultra massive MIMO channels, where a geometrical model is discussed to capture the effects of spatial non-stationarities. The discussed model is based on the massive MIMO extension of the COST 2100 model, which includes the concept of dynamic cluster appearance and disappearance that is unique to both link ends via separable scatterer \emph{visibility regions} \cite{FLOREDELIS1}. In a similar line, a discussion on the implication of IRSs is presented in \cite{BJORNSON1}, where the implications of large-scale fading variability is characterized via first principles. {\color{black}From a measurement perspective, the major limitation of characterizing propagation channels of such large dimensions is the extended measurement \emph{run time} (true for switched and/or virtual arrays), during which the channel is assumed to remain quasi-static. Typically, it is expected that one measurement will take on the order of tens-of-minutes or longer (depending on the measurement bandwidth), limiting the potential measurement scenarios. Fully parallel measurements are not foreseen due to the high cost of up/down-conversion chains and net energy consumption.}

\vspace{-11pt}
\subsection{Propagation in Industrial Environments}
\vspace{-1pt}
Tremendous progress is observed in understanding the nature of wave propagation in industrial environments at both sub-6 GHz and mmWave frequencies (see e.g.,  \cite{CHEFFENA1,SCHMIEDER1,ALSAADEH1,JAECKEL1} for a taxonomy). Naturally, the typical industrial environment is unlike the residential or other indoor environments, since the effects of mechanical and electrical noise, as well as interference are high due to the broad operating temperatures, heavy machinery and ignition systems  \cite{SEXTON1,CHEFFENA1,SCHMIEDER1,ALSAADEH1}.  Generally, industrial buildings are taller than ordinary office buildings and are sectioned into several working areas, between which there usually exist straight aisles for transportation of materials or for human traffic. Modern factories usually have perimeter walls made of precise concrete or steel material. The ceilings are often supported by metal trusses. Most industrial buildings have concrete floors that can support vehicles and heavy machinery. The object type, size, density, and distribution within a specific environment varies significantly across different environments, playing an important role in characterizing the channel \cite{CHEFFENA1}. The presence of random/periodic movements of workers, automated guided vehicles (AGVs) in the form of robots or trucks, overhead cranes, suspended equipment, or other objects will cause time-varying channel conditions. 

A number of propagation measurements and models in various industrial settings have been conducted. The authors in \cite{JAECKEL1} characterize the large-scale parameters of the industrial channel at 2.37 and 5.4 GHz at the Siemens factory in Neumberg, Germany. In both LOS and NLOS conditions, the shadow fading decorrelation distance was approximately 15 m and 30 m - much larger than the corresponding values of 6 m and 10 m in the standardized 3GPP model \cite{3GPP1}. The azimuth and elevation AOD and AOA spreads did not show much difference relative to the 3GPP model. The study in \cite{TRABL1} proposes a double-directional model with parameters that are tailored at 5 GHz from measured data. A detailed comparison between propagation characteristics at 3.7 GHz and 28 GHz is presented over a bandwidth of 2 GHz in \cite{SCHMIEDER1}, where LOS and NLOS pathloss exponents different to those seen in \cite{JAECKEL1} are reported due to the environmental differences. No substantial difference in the delay spread is seen across the two bands of 3.7 GHz to 28 GHz. At 28GHz, AOA information was extracted and angular power profiles and RMS angular spread were evaluated showing an almost uniformly distributed AOA distribution in NLOS conditions across 360$^{\circ}$. The characterized parameters agree with those standardized by the 3GPP. {\color{black}Many further investigations are required to understand the time-varying nature of industrial channels at both below 6 GHz and mmWave frequencies, where not many results exist. For further discussions, the reader is referred to \cite{SEXTON1,CHEFFENA1,SCHMIEDER1,ALSAADEH1,TRABL1,3GPP1}}. 

\vspace{-13pt}
\subsection{UAV Propagation Channels}
\label{UAVPropagationChannels}
UAVs include small drones flying below the regular airspace - low altitude platforms, drones in the regular airspace and high altitude platforms in the stratosphere. Depending on how and where they are operated, the channel properties naturally differ \cite{Khawaja19}. In all cases, one should distinguish the \emph{Air-to-Ground (AG)} channel and the \emph{Air-to-Air (AA)} channel. There are a number of recent survey papers for UAV operation below 6 GHz at low altitudes, see e.g., \cite{Khawaja19,Yan19}. 
Typically, the AA channel behaves as a free-space channel with very limited scattering and fading \cite{Khawaja19}. Given proper alignment, the use of higher frequencies and even free-space optics  are well supported \cite{Dabiri18}. For the AG channel, there is typically more scattering in general, especially at lower frequencies. Often, reflection at the dielectric half-space is strong, giving rise to a two-path fluctuating behavior of the channel. For ground stations located close to the ground level, shadow fading arises as a major limitation, especially at mmWave and above frequencies. Small-scale fading in AG channels usually follows the Ricean distribution with 
$K$-factors in excess of 12 dB. The AG channel can exhibit significant rates of change, with higher order Doppler shifts. In addition to the path loss, the airframe of the UAV can introduce significant shadowing, when the body of the aircraft may obstruct the LOS path. 

The 3GPP has a study of LTE support for UAVs \cite{3GPPuav}. Here a channel model is provided for system-level simulations catering to three environments: rural macrocell, urban macrocell, and urban microcell, respectively. For mmWave UAV channels, the literature is more scarce,  especially with respect to empirical studies. The authors in \cite{Semkin17} analyze 60 GHz UAV-based communication with ray-tracing approach where a detailed description of the environment is achieved by a photogrammetric approach. With an accurate and detailed description of the environment and proper calibration, ray-tracing methods are able to provide accurate predictions of the expected channel behavior in this use case \cite{Semkin17}. UAVs are also explored to provide cellular coverage in remote areas via high altitude platforms. The authors of \cite{Cao19} gives an overview of propagation properties of high altitude platforms. In June 2020, Loon and Telkom in Kenya launched their first commercial service providing 4G services from a set of balloons circling in the stratosphere at an approximate altitude of 20 km. This is in stark contrast to LEO or  geostationary satellites operating from altitudes of 300-1200 km and 36000 km, respectively. This is important because of the \emph{latency} induced. The propagation delay for two-way communication is in the order of 0.1 ms rather than in the 2-8 ms range for LEO satellites or 240 ms for geostationary satellites. To this end, such platforms have the possibility to support real-time services with tight latency requirements.

\vspace{-11pt}
\subsection{Vehicular Propagation Channels}
\label{VehicularPropagationChannels}
The behavior of V2V and V2I channels below 6 GHz is well investigated and understood. The authors of \cite{Mecklenbrauker11} give an overview of important characteristics and considerations for sub-6 GHz V2V communication. Six important propagation characteristics are: 1) The channel cannot be seen as wide sense stationary with uncorrelated scattering; the statistics both in terms of time correlation and frequency correlation change over time \cite{Bernado14}. 2) High Doppler spreads may occur due to the high relative movements from transmitter to the receiver. In certain cases, up to 4$\times$ higher Doppler spread is experienced compared to a conventional cellular scenario with a stationary BS. 3) In a highway scenario, the channel is often \emph{sparse} with a few dominant MPCs. V2V channels in urban scenarios tend to be much richer in its multipath structure  \cite{Gustafson20}. 4) MPCs (especially in urban settings) tend to have a limited lifetime with frequent deaths and births \cite{Mahler17}. 5) Blocking of the LOS by other vehicles tend to have significant impact of the path loss. The median loss by an obstructing truck was reported to be 12-13 dB in \cite{Vlastaras14}. 6) The influence of the antenna position and antenna pattern should not be underestimated \cite{Mecklenbrauker11}. They affect not only the path loss, but also the statistics of the channel parameters. 

When going up in frequency, it can be expected that those properties not only remain, but become even more exaggerated. The authors of \cite{Boban19}
give an up-to-date overview of mmWave V2V channel properties. It is noteworthy that there is a lack of measurement results for  mmWave vehicular channels, and most conclusions are drawn from stationary measurements. For both below and above 6 GHz, 3GPP TR 37.885 \cite{3GPP_37_385} presents a standardized V2V channel model for system simulations, that is based on the tapped delay line principle. Above 6 GHz, it is assumed that the simulated bandwidth is 200 MHz with an aggregated bandwidth of up to 1 GHz. For 6G, one of the main use cases is cooperative perception, where raw sensor data from, e.g., camera and radars is shared between vehicles. The anticipated data rates for such applications are up to 1 Gbps calling for use of the wider bandwidths available at mmWave frequencies. One of the few dynamic mmWave measurement campaign for a V2I scenario is presented in \cite{Park19}. For a highway scenario, with vehicle mobility of 100 km/h, the Doppler spread experienced for a carrier frequency of 28 GHz was up to 10 kHz. As a rough estimate, this gives a worst-case coherence time as low as 100 $\mu$s, which is extremely small for conventional pilot-based OFDM transmission. The study in \cite{Groll19} analyzed the sparsity of the 60 GHz V2I channel. It was concluded that the sparsity in the delay-Doppler domain holds true also in the measured urban street crossing scenario, and that a single cluster with a specific delay Doppler characteristics was dominating, hence enabling compensation of the delay and Doppler shifts and being suitable for OTFS type of modulation. The authors of \cite{Kampert18} analyzed the influence of a realistic antenna mount near the vehicle headlights. The measured antenna pattern showed similar irregularities as seen at sub-6 GHz, with excess path loss typically ranging from 10 to 25 dB depending on the AOA, and more pronounced variations from 74-84 GHz in contrast to 26-33 GHz. In \cite{Boban19}, the influence of LOS was discussed. With directional antennas, the channel can be modelled with two-paths at the measured frequencies of 38, 60 and 76 GHz. Blocking the LOS results in excess losses in the range of 5-30 dB depending on the particular scenario and frequency, i.e., in the \emph{same} range as reported for sub-6 GHz V2V communication. The blockage of the LOS also results in sudden increases in the angular spread and delay spread, again affecting the channel statistics. For other types of channels, in particular the ones experienced in railway systems, we refer the reader to discussions in \cite{Ai_et_al_2020_ProcIEEE}.

\vspace{-12pt}
\subsection{Wearable Propagation Channels}
\label{WearablePropagationChannels}
Wearable devices are important in healthcare systems, robotics, immersive video applications. Thus far, there are no standardized models for body area networks, though many studies are reported, see e.g., \cite{hall,5272227}.
The existing measurements can be categorised as narrowband for $300$ kHz-$1$ MHz at sub-1 and 2 GHz frequencies. In contrast, there also exist ultra wideband measurements with a  measurement bandwidth of 499 MHz in the C-band and 6-10 GHz. Here one of the most extensive studies is by the authors in \cite{sangodoyin2018impact}, which takes into account 60 human subjects. Models for large-scale and small-scale fading are provided, yet the models given are specific to the measured body locations (i.e., where the sensors are placed), antenna types, frequency bands, proving difficult to generalize to other bands and locations. {\color{black}This seems to be a major challenge requiring much further work.}

Continuing the top-down look at 6G systems, the following section evaluates the design challenges in real-time signal processing and RF front-end architectures, as well as describes possible solutions to realize working systems across a wide range of frequencies. The section begins with a discussion on the implications of increasing carrier frequencies. 

\vspace{-9pt}
\section{Real-Time Processing and RF Transceiver Design: Challenges, Possibilities, and Solutions}
\label{transceiverdesign}
\subsection{Implications of Increasing Carrier Bandwidths}
\vspace{-1pt}
While the operating bandwidths of some of the windows in Tab.~\ref{table:2} spans tens of GHz, building a radio with a single carrier over the entire bandwidth is almost impossible, especially if one wants to maintain equally high performance and energy efficiency across the band by retaining the linearity of RF front-end circuits. In recognition of this, even for 5G systems in case of mmWave bands, the maximum permissible carrier bandwidth is 400 MHz. On a similar line, close proximity services even in the THz bands are being considered to be given a maximum bandwidth of 1 GHz \cite{M2417}. This is rather astonishing, since in the first place, the adoption to mmWave and THz frequency bands was driven by the fact that orders-of-magnitude more bandwidths could be leveraged relative to canonical systems. Current commercial equipment at mmWave frequencies is made up of aggregating 4 carriers, each 100 MHz wide. However, the maximum carrier bandwidth for mmWave systems defined in 3GPP is 400 MHz. Relative to a 100 MHz carrier, the \emph{noise floor} of a receiver using 1 GHz bandwidth will be 10 dB higher, causing SNR degradation by 10 dB. As such, in practice, the bandwidth of a single carrier could  be limited to 100 MHz, yet higher bandwidths can be obtained by aggregating component carriers. Following this line of thought, if 10 GHz bandwidth is desired, one has to aggregate 100 such carriers. {\color{black} A direct consequence of this is that the radio hardware has to be in \emph{calibration} across the 100 carriers - something which poses a tremendous challenge at such high frequencies, particularly as the effects of phase noise start to dominate. With such wide bandwidths, the radio performance at the lower end of the band can be expected to be entirely different from the upper end of the band. To this end, the maximum number of carriers, and in turn the maximum operable bandwidth, will be a compromise based on the ability to obtain antenna integrated RF circuits and effective isotropic radiated power limits for safety. We note that this is a significant design challenge.}

\vspace{-13pt}
\subsection{Processing Aspects for mmWave and THz Frequency Bands}
\label{ProcessingAspectsforTerahertzFrequencyBands}
It is clear that the high electromagnetic losses in the THz frequency bands pose a tremendous research and engineering challenge. Realistically, it is difficult to imagine (some) 6G services beyond window W1, between 140-350 GHz in Fig.~\ref{AbspPeaks}. Here, the free-space loss at a nominal link  distance of 10 m is well in excess of 100 dB.\footnote{In the context of the immediate future, extension of 5G operations up to 71 GHz is already under consideration in 3GPP for Release 17. We envisage this trend to continue beyond 100 GHz, leading into 6G systems.} A direct consequence of this is limited cell range - a trend which is emerging from 5G systems from network densification. To overcome this issue, the proposal of ultra massive MIMO systems has been made in the THz literature, which is envisaged to close the link budget by integrating a very large number of elements in minuscule footprints to increase the link distance. This is critical for the earlier mentioned 6G use cases requiring Tbps connectivity. Ultimately, the energy consumption along with the exact type of beamforming architecture will put a practical constraint on the realizable number of elements which are considered at the BS and UE link ends. 

To meet the target of up to Tbps connectivity, three-dimensional spatial beamforming will be critical. The complete three-dimensional nature of the propagation channel is \emph{not} utilized even in 5G systems at mmWave frequencies, where \emph{analog beamforming} is mostly implemented in commercial products with multiple antenna panels (with or without shared fronthauling), each being able to form one beam towards a pre-defined direction. On the other hand, progress in RF circuits has been tremendous to realize radio transceivers with \emph{fully digital beamforming} for bands below 6 GHz, and more recently at mmWave bands from 24.5-29.5 GHz \cite{DBF2}. Nevertheless, implementing fully digital beamforming at THz frequencies is a formidable task, with an order-of-magnitude higher complexity relative to mmWave bands. It should not be taken for granted that in a ``matter of time", RF electronics will mature, and we will be able realize digital beamforming even at THz. \emph{As for 5G systems, for the short-to-medium term, phased array implementations performing analog or hybrid beamforming seem most likely.} Unlike for microwave and mmWave frequencies, for the THz bands, the phased array processing architecture needs to be redesigned due to the complexities in antenna fabrication, high speed/high power mixed signal components, RF interconnects and heat dissipation. The most common type of antenna implementation in microstrip patch elements do not operate efficiently at THz frequencies due to the high dielectric and conductor losses at the RF substrate level. As such, phased arrays fabricated with \emph{nano materials}, such as graphene have been extensively discussed to build miniature plasmonic antennas with dynamic operational modes to reap the benefits of spatial multiplexing and beamforming \cite{akyildizdistance}. On the other hand, \emph{metamaterial}-based antennas, hypersurfaces, and RF front-end solutions are also emerging as a key technology \cite{SAMSUNGWP1}. To increase the beamforming gain, the concept of \emph{metasurface lenses} is introduced, which acts as a RF power splitting, phase shifting, and power combining network that is applied to the radiated signal from an antenna array \cite{SAMSUNGWP1}. Such a structure has the potential to replace conventional RF power splitting, phase shifting and power combining circuits, which are complex and power hungry, with a relative cheap passive device (in the form of a lens), yielding significant gains in circuit complexity and energy consumption \cite{AB3}. A more detailed discussion about such technologies is given in \cite{akyildizdistance,SAMSUNGWP1}. 

{\color{black}From a real-time processing viewpoint, the major challenge at both mmWave and THz frequencies is in the \emph{dynamic control} and \emph{management of RF interconnects} of the array elements and the associated beamforming networks. While this problem was present in the mmWave bands, the challenge is elevated even higher due to the even shorter channel coherence times (for a fixed Doppler spread), higher phase noise, and higher number of antenna elements. Even with hybrid beamforming, to manage the processing complexity as well as the cost, \emph{fully-connected} architectures which require dedicated phase shifters per-RF signal path will be cost prohibitive - and a design based on the array of sub-arrays principle must be leveraged \cite{SAMSUNGWP1,akyildizdistance}. Here, a subset of antennas are accessible to one specific RF chain, while at baseband, a digital processing module is implemented for both structures to control the data streams and manage interference among users. Low-resolution ADCs and DACs must also be exploited to manage the cost and implementation of transceivers. For THz bands, further discussion is given in the following subsection.}

To assess when it may be likely for us to achieve Tbps rates, we carry out a toy example. For the sake of argument, we assume perfect CSI and ideal transceiver architectures at both the BS and UE sides, where 4096 elements are employed at the BS, and 16 elements are employed at the UE, both in uniform planar arrays (UPAs) of 64$\times$64 and 4$\times$4 elements, respectively. For both UPAs, the horizontal spacing was set 0.5$\lambda$, while the vertical spacing was 0.7$\lambda$, with an example per-element pattern from \cite{3GPP1}. The antennas were driven across two separate bandwidths: 140-141 GHz and 140-240 GHz across a link distance of 15 m. For both bandwidths, the noise floors computed using the classical noise floor expressions. The propagation channel impulse responses were obtained from the model in \cite{PRIEBE1}. Figure~\ref{THzCapacityPriebe} demonstrates the single-user MIMO capacity cumulative distribution functions (CDFs) at SNR=10 dB and SNR=3 dB. As seen from the top subfigure, with bandwidth of 100 GHz at 10 dB SNR, peak capacity of 1 Tbps can be achievable in theory (indicated on the figure with a green diamond) under the assumptions mentioned above. An almost constant loss in capacity is observable across all CDF values when the operating SNR is reduced from 10 dB to 3 dB. A comparison of the same SNR levels with a bandwidth of 1 GHz yields \emph{less than} a 100$\times$ capacity difference due to bandwidth appearing in the pre-log factor of the capacity formulation. It is noteworthy that the bandwidth term plays a much more prominent role in the capacity predictions, in contrast to the improved SNR (which features inside the logarithm) due to lower noise floor at 1 GHz relative to 100 GHz. {\color{black} With this in mind, one can readily ask many questions about how such high capacities can be achievable under realistic CSI and transceiver architecture constraints, despite the aforementioned difficulties in real-time operation. If we would like to operate a system on a common constellation, is it practically feasible to achieve forward link SNRs on the order of 10 dB? Would the modulation and coding gains be able to maintain such high SNRs for a long time period? Large bandwidths are indeed available at THz frequencies, however are we able to utilize these bandwidths with realizable beamforming architectures? These are all major research questions that need to be answered.}
\begin{figure}[!t]
    \vspace{-10pt}
    \centering    
    \hspace{-15pt}
    \includegraphics[width=8cm]{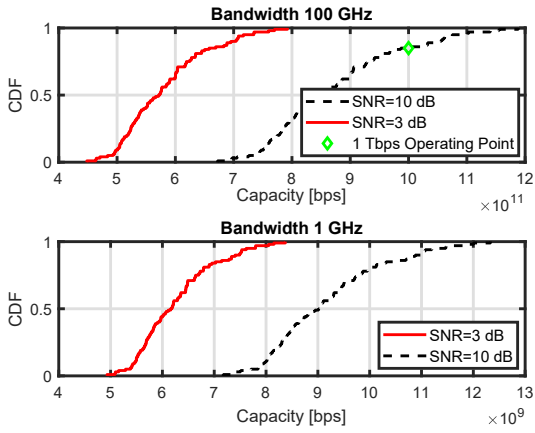}
    \vspace{-12pt}
    \caption{Single-user MIMO capacity CDFs with 4096 BS antennas serving a UE with 16 antennas over 1 GHz and 100 GHz bandwidths. The impulse responses were generated from \cite{PRIEBE1}.}
    \label{THzCapacityPriebe}
    \vspace{-14pt}
\end{figure} 

In the context of multiuser systems, as a simple approximation, the per-UE capacity, $R$, can be thought of as 
\vspace{-1pt}
\begin{equation}
\label{peruserrate}
R \approx \left(\frac{BL}{K}\right) \textrm{SE}. 
\vspace{-1pt}
\end{equation}
where $B$ and $L$ are the bandwidth and number of MIMO layers for a total of $K$ UEs, and SE is the instantaneous spectral efficiency given by $\textrm{SE}\approx\log_{2}\left(1+\textrm{SINR}\right)$, where $\textrm{SINR}$ denotes the signal-to-interference-plus-noise ratio of a given UE. Now to increase the capacity, we need to increase $B$, $L$ and the SINR \cite{tatariathesis}. Increasing $B$ is certainly possible in the THz bands, yet power density decreases with increasing bandwidth. Increasing MIMO layers will need ultra massive MIMO arrays at both ends, yet they can only be exploited fully if the propagation channel can support a reasonable rank - something which is largely unknown from the sparsely explored THz literature (except for studies such as \cite{PRIEBE1}). Ultra high dimensional arrays will result in extreme directivity in transmitted beams, which will reduce interference. Yet the increasing bandwidth will also increase the noise floor (as mentioned previously). Finally, network densification will decrease the number of competing users $K$, yet this will also increase the network operational expenditure and BS coordination overheads. {\color{black}Going forward, all of these factors must be carefully considered in the context of THz research.}
\begin{figure*}[!t]
    \centering
    \hspace{20pt}
    \includegraphics[width=10.8cm]{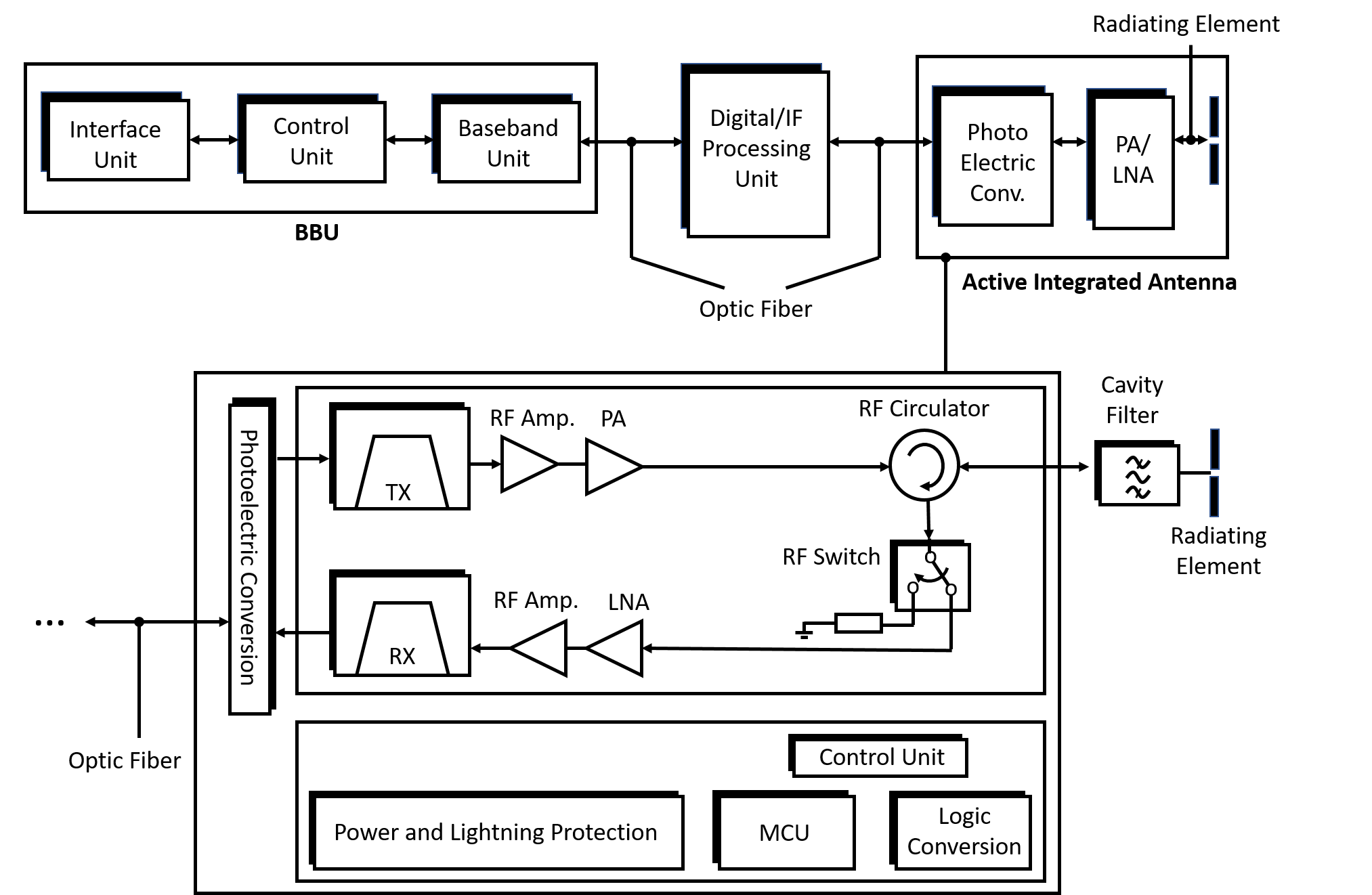}
    \vspace{1pt}
    \caption{Illustration of a typical BS transceiver architecture for sub-6 GHz and mmWave frequencies with radio-over-fiber and active integrated antenna elements. In order to avoid ambiguity, only one radiating element is shown. The figure is reproduced from \cite{ITURM23340}. The terms IF, PA, LNA and MCU denote intermediate frequency, power amplifier, low-noise amplifier, and microcontroller unit, respectively.}
    \label{AIRT}
    \vspace{-10pt}
\end{figure*}

\vspace{-9pt}
\subsection{RF Transceiver Challenges and Possibilities}
\label{RFTransceiverChallengesandPossibilities}
For sub-6 GHz and mmWave frequencies, a typical BS transceiver architecture is depicted in Fig.~\ref{AIRT} \cite{ITURM23340}, where an \emph{amalgamation} of radio-over-fiber and active integrated antennas are utilized. In order to avoid cluttering the figure only one radiating element is demonstrated. The up-conversion and down-conversion processes are controlled in real-time via the depicted control modules and the RF circulator. The transmitter and receiver, denoted as TX and RX in the fig. perform the mixing and de-mixing operations. For transmission and reception, a two-stage cascaded amplifier sequence is used to provide additional power gain. Additional filtering and control circuits which are critical to the transceiver operation are also demonstrated. {\color{black} While such architectures can be realized at sub-6 GHz and mmWave frequencies thanks to the progress in RF circuits, the same cannot be said for the THz bands. Using the THz band will impose major challenges on the transceiver hardware design. First and foremost, operating at such high frequencies puts stringent requirements on the semiconductor technology. Even when using state-of-the-art technology, the frequency of operation will approach, or in extreme cases even exceed, the frequency, $f_{\textrm{max}}$, where the semiconductor is able to successfully provide a power gain. The achievable receiver noise figure as well as transmitter efficiency will then be severely degraded compared to operation at lower frequencies. To maximize the high frequency gain, the technology must use scaled down feature sizes, requiring low supply voltage to achieve reliability, reducing the achievable transmitter output power. Combined with the degraded receiver noise figure, the reduced antenna aperture, and the wide signal bandwidth will naturally result in very short link distances, unless  ultra massive number of elements are combined coherently with sharp beamforming. Thousands to tens-of-thousands of antenna elements may be required for THz BSs.}

For the sake of example, operating at 500 GHz with ten thousand antenna elements brings the size of the required array down to just 3 cm $\times$ 3 cm, with the elements spaced half wavelength apart, i.e., 0.3 mm. {\color{black}\emph{The RF electronics must have the same size, to minimize the length of THz interconnects, which is a major research challenge.} Each chip must then feature multiple transceivers. For instance, a 3 mm $\times$ 3 mm chip can have 100 transceivers, and 100 such chips need to be used in the ten-thousand antenna element array. The antennas may be implemented on or off chip, where on chip antennas generally have less efficiency, yet they eliminate the loss in chip-to-carrier interfaces. In addition, heat dissipation becomes a major problem. Since THz transceivers will have low efficiency, the area for heat dissipation will be very small. \emph{If each transceiver consumes 100 mW, the total power consumption of the array becomes 1 kW, having major implications on the system not being able to be continuously active.} If heat dissipation becomes too problematic, more sparse arrays may have to be considered, for e.g., using compressive sensing-based \emph{array thinning} principles with more than half wavelength element spacing \cite{LECCI1}. However, this would cause side lobes that need to be managed which in turn may pose constraints on spectrum sharing with existing or adjacent services.}

To create e.g., 10000 transceivers with high level of integration, a silicon-based technology must be used. While silicon metal oxide semiconductor field effect transistor (MOSFET) transistors are predicted to have reached their peak speed, and will actually \emph{degrade} with further scaling, silicon germanium (SiGe) bipolar transistors are predicted to reach an $f_{\textrm{max}}$ of close to 2 THz within a 5 nm device  \cite{Schroter}. In such a technology, amplifiers and oscillators up to about 1 THz could be realized with high performance and integration. With today's silicon technology, however, 500 GHz amplifiers and oscillators cannot be realized, and to operate at such frequencies, frequency multiplication in a non-linear fashion is necessary. A transmitter based on a frequency multiplier, or a receiver with a sub-harmonic mixer, however, will not reach attractive performance. Currently, a better option may then be to use indium phosphide (InP) technology for the highest frequency parts, combined with a silicon complementary metal–oxide–semiconductor driven baseband circuit. Amplifiers and mixers at 800 GHz have been demonstrated in 25 nm InP high-electron-mobility transistor (HEMT) technology with an $f_{\textrm{max}}$ of 1.5 THz \cite{Leong}. When 5 nm SiGe technology becomes available, the level of integration will be higher, resulting in reduced production costs. {\color{black}\emph{We believe this to be a must for implementations of ultra massive MIMO arrays.}}

{\color{black}Another important challenge is the generation of coherent and low noise local oscillator (LO) signals for ten thousand or more transceivers. The generation of a central 500 GHz signal to be distributed to all transceivers, perhaps 100, on a chip seems impractical, as it would consume very large power in the buffers. As such, a more distributed solution with \emph{local} phase locked loops (PLLs) is more appealing, since a lower frequency reference can then be distributed over the chip. The phase noise of different PLLs will then be non-correlated; using multiple PLL signals together can achieve low noise beams . On the other hand, doing this results in depth reduction when forming notches, limiting the performance of multiple simultaneous beams \cite{LaCaille}. To this end, there is a trade-off in choosing the number of PLLs. Nonetheless, given the high power of LO signal distribution, a large number of PLLs seems favorable. This is further pronounced by the difficulty of reaching high resonator energy in a single oscillator at such high frequencies, making it attractive to increase the total energy by increasing the number of oscillators in the system. Using a large number of PLLs also provides LO beamforming possibilities, as the PLL phase can accurately be controlled \cite{Axholt}. Regardless of LO architecture, another challenge is frequency tuning of oscillators, since the quality factor of variable reactances (varactors) is inversely proportional to the operating frequency. As such, at THz frequencies, other tuning mechanisms should be investigated, like using resistance for tuning \cite{Huang}. All of these challenges call for substantial research efforts in this important direction, and must be overcome to realize systems that are envisioned for 6G networks.}

\vspace{-13pt}
{\color{black}\subsection{Comments on Energy Consumption and Efficiency}
\label{CommentsonEnergyConsumptionandEfficiency}
\vspace{-3pt}
As the amount of data to communicate and process is increased by orders-of-magnitude, energy efficiency becomes critical, especially in battery powered devices. The increased antenna gain from using large arrays at high mmWave and early terahertz bands will help energy efficiency by directing the transmitted energy towards the desired UEs, and so will reduced cell sizes, required to meet the high peak rates. At the same time PA efficiency and UE noise figure will degrade with frequency, counteracting some of the gains of using more directed transmissions over shorter range. With advances in semiconductor technology, however, like scaled SiGe bipolar technology, the noise figure and power efficiency is 
predicted to become attractive even at these frequencies \cite{DE1}. The power consumption of a large array transceiver may still be high, due to the many transceiver up/down-conversion channels, but the data rate can be extremely high and the energy per-bit is expected to 
drop by orders-of-magnitude compared to existing cellular systems. The bottleneck for energy efficiency may then become processing the data, e.g., to display a hologram from an extremely high data rate stream. According to Gene's law for baseband processing, similar trends to those mentioned above have been observed \cite{DE1}. However, in the last five years, they have shown signs of slowing down substantially to 10$\times$ improvements per-decade, and has been outpaced by the 12$\times$ improvement per-decade of GPUs. For the coming decade, pure technology scaling will only bring up to 4$\times$ energy reduction considering the small number of upcoming CMOS generations \cite{DE1}. To this end, technology scaling needs to be supplemented with significant and coordinated advances at all levels of abstraction. Such considerations also hold for voltage scaling, which has been extensively leveraged in the last two decades. Specifically, 0.6 V operation is already available for commercial processors and standard cell libraries, which is rapidly approaching the transistor threshold and hence leaving limited opportunity for further scaling \cite{DRESLINKSKI1}. In the same line, parallelism is no longer providing the energy savings that it used to, especially for high speed applications whose workload may not be naturally parallelizable. For example, the number of simultaneously active cores in state-of-the-art UE platforms is well known to have remained essentially constant in the last two generations, and hence management of UE power consumption needs to be more dependent on network-side energy saving mechanisms. Looking ahead towards the next decade, novel design dimensions will be needed to trade-off energy consumption and reduce it, whenever the related specifications can be relaxed.}

\vspace{-11pt}
\section{Conclusions}
\label{conclusions}
\vspace{-2pt}
To the best of our knowledge, this paper is the first to take a holistic top-down approach in describing 6G systems. The paper begins by presenting a vision for 6G, followed by a detailed breakdown of the next generation use cases, such as high fidelity holographic communications, immersive reality, tactile internet, vastly interconnected society and space-integrated communications. For each use case, we present a breakdown of its technical requirements. This is followed by a discussion on the potential deployment scenarios which 6G systems will likely operate in. A rigorous discussion of the research challenges and possible solutions that must be addressed from applications, to design of the next generation core networks, down to PHY is presented. Unlike other studies, we differentiate between what is theoretically possible, and what may be practically achievable for each aspect of the system. {\color{black}In the deployment of 6G systems, backwards compatibility must be considered. This is since devices will be multimode and multiband. A 6G device will need to fall back to 5G and 4G depending upon the coverage conditions. Therefore, the 6G RAN and core network must be backwards compatible with the previous generations. There will be significant challenges and design trade-offs to achieve this; e.g., the introduction of a new network architecture for the 6G core network, as discussed in the paper. This also applies to waveform and coding methods, where a large number of them will not be backwards compatible to what is introduced in 5G.} After a lengthy analysis dissecting many system components, as well as exploring possible solutions, we can conclude that there is an exciting future that lies ahead. The road to overcome the challenges is full of obstacles, yet we provide enough insights to begin research towards promising  directions. This will serve as a motivation for research approaching the next decade.}

\vspace{-11pt}
\section*{Acknowledgments}
\vspace{-2pt}
The authors would like to express their sincere appreciation to the anonymous reviewers whose suggestions have significantly improved the manuscript. The authors also express their sincere gratitude to Prof. Catherine Rosenberg, for her constructive feedback and comments, which have also contributed to considerable improvements of this manuscript.

\vspace{-15pt}
\bibliographystyle{ieeetr}
\bibliography{main.bib}

\begin{IEEEbiography}[{\includegraphics[width=1.05in,height=1.25in,clip,keepaspectratio]{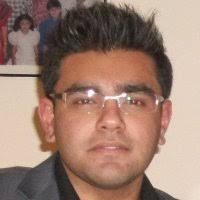}}]{Harsh Tataria} received a B.E. degree in Electronic and Computer Systems Engineering (Hons) and Doctor of Philosophy (Ph.D.) degree in Communications Engineering from Victoria University of Wellington, New Zealand, in December 2013 and March 2017, respectively. Since then, he has held Postdoctoral Fellowship positions at Queen's University Belfast, UK, University of Southern California, US, and Lund University, Sweden. Currently, he is an Assistant Professor of Communications Engineering at Lund University. His research interests include measurement and modelling of propagation channels, multiple antenna transceiver design and statistical analysis techniques of multiple antenna systems at centimeter-wave, millimeter-wave and sub-terahertz frequencies. 
\end{IEEEbiography}

\begin{IEEEbiography}[{\includegraphics[width=1.15in,height=1.25in,clip,keepaspectratio]{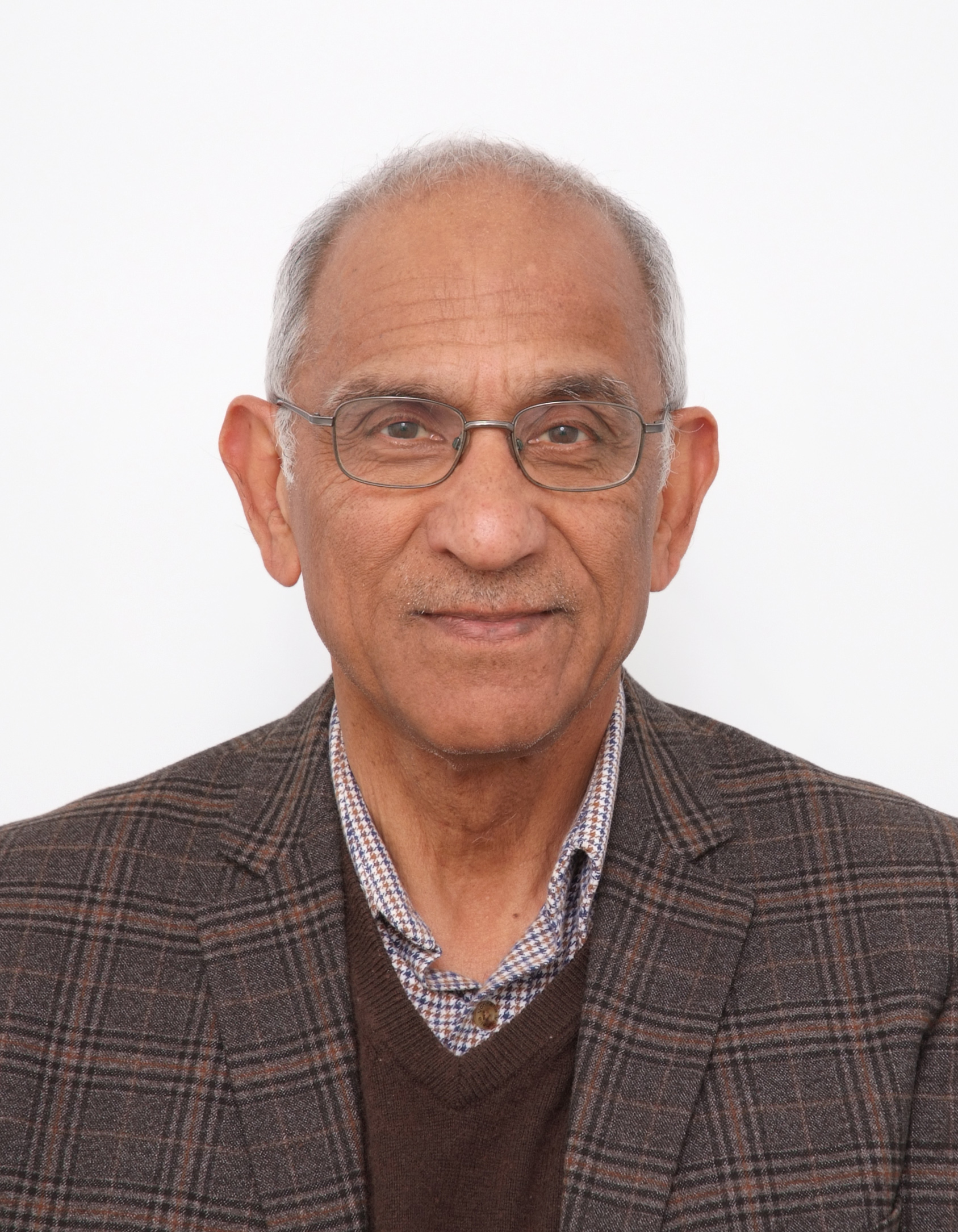}}]{Mansoor Shafi} (S'69–M'82–SM'87–F'93–LF'16) received the B.Sc. (Eng.) and Ph.D. degrees in electrical engineering from the University of Engineering and Technology Lahore and The University of Auckland in 1970 and 1979, respectively. From 1975 to 1979, he was a Junior Lecturer with The University of Auckland, he then joined the New Zealand Post Office, that later evolved to Telecom NZ, and recently to Spark New Zealand. He is currently a Telecom Fellow (Wireless at Spark NZ) and an Adjunct Professor with the Schools of Engineering, Victoria University and Canterbury University, respectively, in New Zealand.  He is a Delegate of NZ to the meetings of ITU-R and APT and has contributed to a large number of wireless communications standards. His research interests include radio propagation, the design and performance analysis for wireless communication systems, especially antenna arrays, MIMO, cognitive radio, and massive MIMO and mmWave systems. He has authored over 200 papers in these areas. He has co- shared two IEEE prize winning papers: the IEEE Communications Society, Best Tutorial Paper Award, 2004 (co-shared with D. Gesbert, D.-S. Shiu, A. Naguib, and P. Smith) for the paper, From Theory to Practice: An overview of MIMO Space Time Coded Wireless Systems, IEEE JSAC, April 2003, and the IEEE Donald G Fink Award 2011, (co shared with A. Molisch and L. J. Greenstein), for their paper in IEEE Proceedings April 2009, Propagation Issues for Cognitive Radio. 
Dr. Shafi has also received the IEEE Communications Society Public Service Award, 1992 “For Leadership in the Development of Telecommunications in Pakistan and Other Developing Countries,” and was made a member of the New Zealand Order of Merit, Queens Birthday Honors 2013, For Services to Wireless Communications. He has been a Co-Guest Editor for three previous JSAC editions, the IEEE Proceedings, and the IEEE Communications Magazine, and a Co-Chair of ICC 2005 Wireless Communications Symposium, and has held various editorial and TPC roles in the IEEE journals and conferences.
\end{IEEEbiography}

\begin{IEEEbiography}[{\includegraphics[width=1.05in,height=1.25in,clip,keepaspectratio]{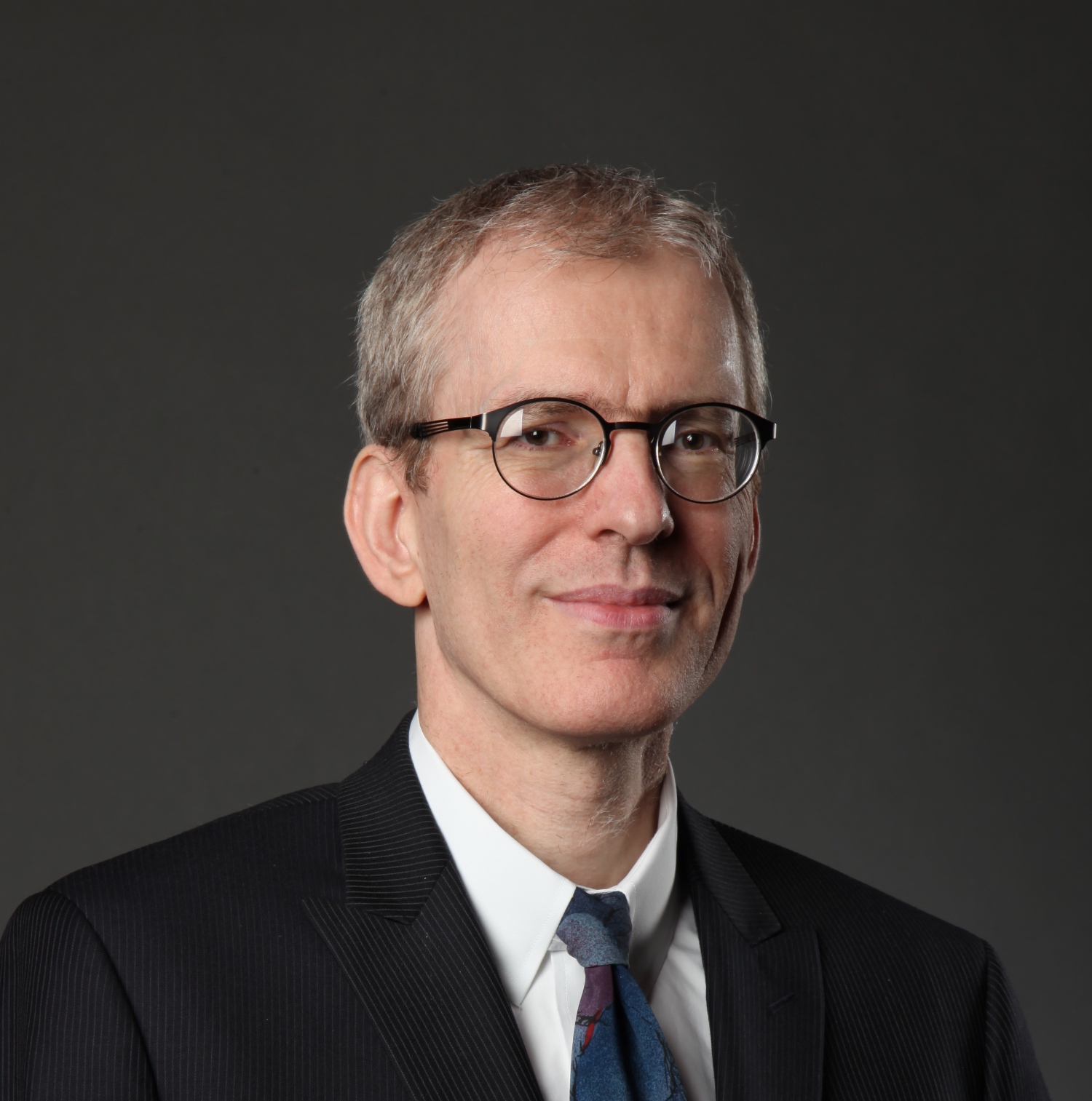}}]{Andreas F. Molisch} received his degrees (Dipl.Ing. 1990, PhD 1994, Habilitation 1999) from the Technical University Vienna, Austria. He spent the next 10 years in industry, at FTW, AT\&T (Bell) Laboratories, and Mitsubishi Electric Research Labs (where he rose to Chief Wireless Standards Architect). In 2009 he joined the University of Southern California (USC) in Los Angeles, CA, as Professor, and founded the Wireless Devices and Systems (WiDeS) group. In 2017, he was appointed to the Solomon Golomb – Andrew and Erna Viterbi Chair. 

His research interests revolve around wireless propagation channels, wireless systems design, and their interaction. Recently, his main interests have been wireless channel measurement and modeling for 5G and beyond 5G systems, joint communication-caching-computation, hybrid beamforming, UWB/TOA based localization, and novel modulation/multiple access methods. Overall, he has published 4 books (among them the textbook “Wireless Communications”, currently in its second edition), 21 book chapters, 260 journal papers, and 360 conference papers. He is also the inventor of 60 granted (and more than 20 pending) patents, and co-author of some 70 standards contributions. 

Dr. Molisch has been an Editor of a number of journals and special issues, General Chair, Technical Program Committee Chair, or Symposium Chair of multiple international conferences, as well as Chairman of various international standardization groups. He is a Fellow of the National Academy of Inventors, Fellow of the AAAS, Fellow of the IEEE, Fellow of the IET, an IEEE Distinguished Lecturer, and a member of the Austrian Academy of Sciences. He has received numerous awards, among them the IET Achievement Medal, the Technical Achievement Awards of IEEE Vehicular Technology Society (Evans Avant-Garde Award) and the IEEE Communications Society (Edwin Howard Armstrong Award), and the Technical Field Award of the IEEE for Communications, the Eric Sumner Award.
\end{IEEEbiography}

\begin{IEEEbiography}[{\includegraphics[width=1in,height=1.35in,clip,keepaspectratio]{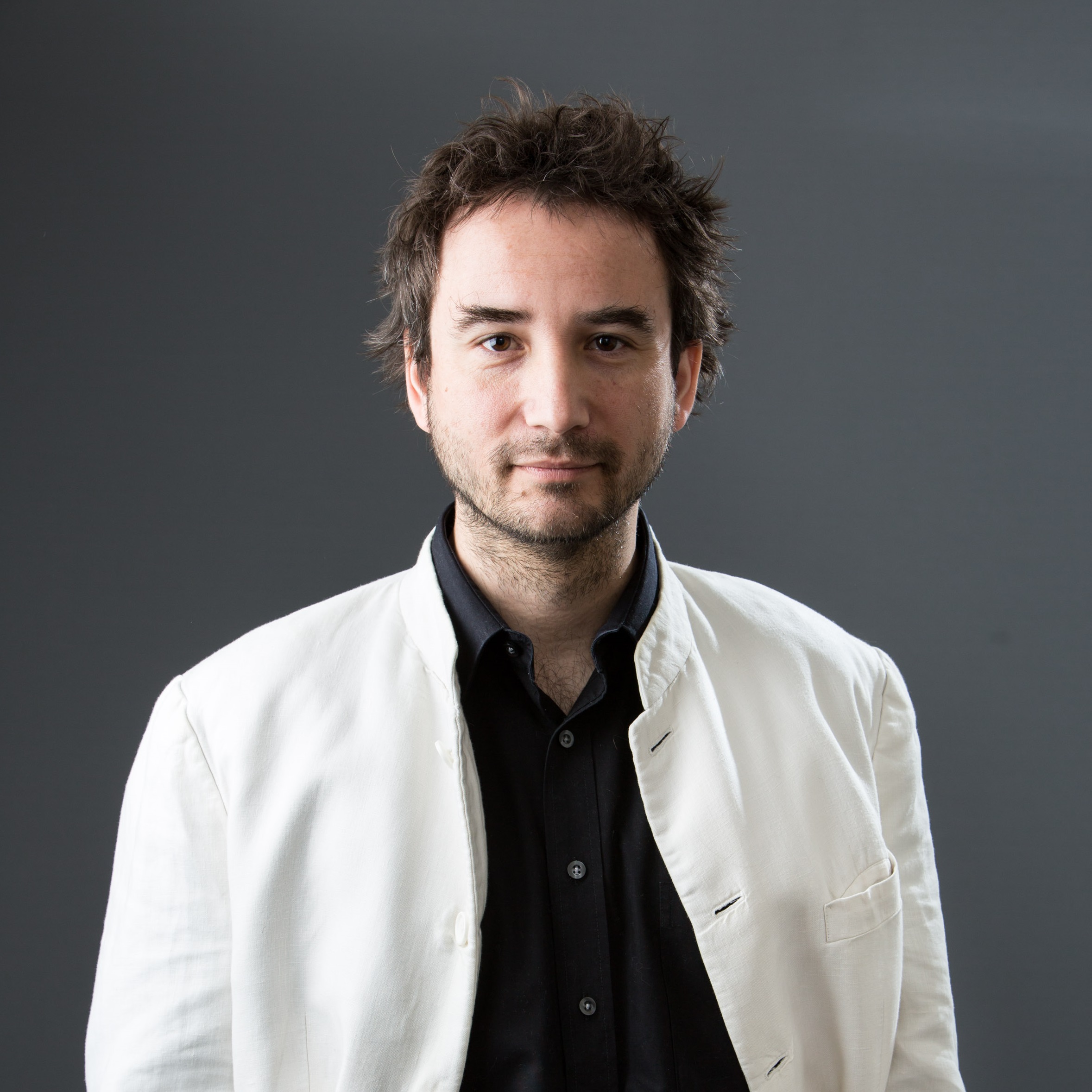}}]{Mischa Dohler} is a full Professor in Wireless Communications at King's College London, driving cross-disciplinary research and innovation in technology, sciences and arts. He is a Fellow of the IEEE, the Royal Academy of Engineering, the Royal Society of Arts (RSA), the Institution of Engineering and Technology (IET); and a Distinguished Member of Harvard Square Leaders Excellence. He is a serial entrepreneur with 5 companies; composer \& pianist with 5 albums on Spotify/iTunes; and fluent in 6 languages. He acts as policy advisor on issues related to digital, skills and education. He has had ample coverage by national and international press and media.
He is a frequent keynote, panel and tutorial speaker, and has received numerous awards. He has pioneered several research fields, contributed to numerous wireless broadband, IoT/M2M and cyber security standards, holds a dozen patents, organized and chaired numerous conferences, was the Editor-in-Chief of two journals, has more than 300 highly-cited publications, and authored several books.
He was the Director of the Centre for Telecommunications Research at King’s from 2014-2018. He is the Cofounder of the Smart Cities pioneering company Worldsensing, where he was the CTO from 2008-2014. He also worked as a Senior Researcher at Orange/France Telecom from 2005-2008.
\end{IEEEbiography}

\begin{IEEEbiography}[{\includegraphics[width=1in,height=1.25in,clip,keepaspectratio]{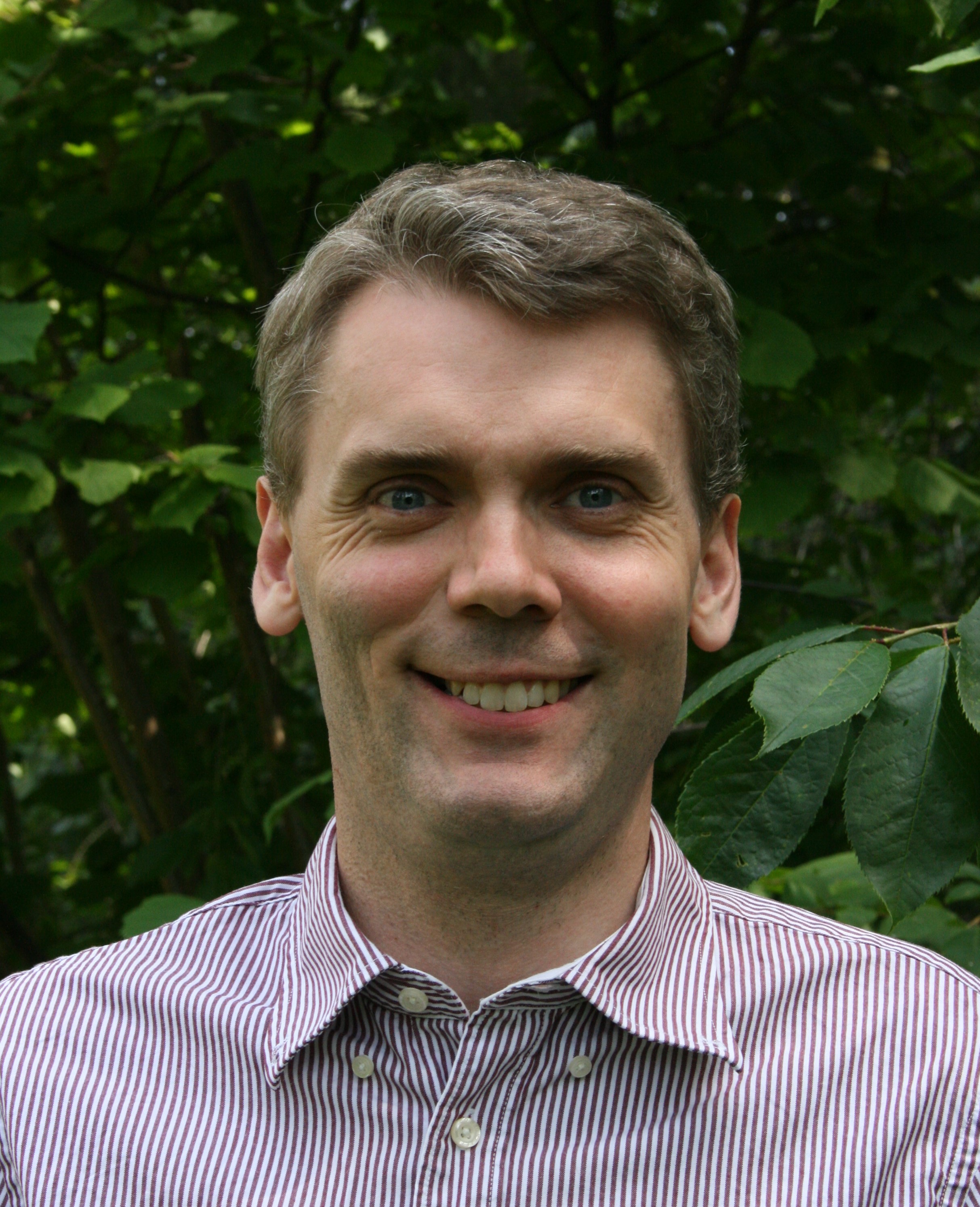}}]{Henrik Sjöland} (M'98-SM'10) received the M.Sc. degree in electrical engineering from Lund University, Sweden, in 1994, and the PhD degree from the same university in 1997. In 1999 he was a postdoc at UCLA on a Fulbright scholarship. He has been an associate professor at Lund University since year 2000, and a full professor since 2008. Since 2002 he is also part time employed at Ericsson Research, where he is currently a Senior Specialist. He has authored or co-authored more than 180 international peer reviewed journal and conference papers and holds patents on more than 30 different inventions. 
Henrik Sjöland is currently an associate editor of IEEE Transactions on Circuits and Systems – I, and he has previously been an associate editor of IEEE Transactions on Circuits and Systems – II and a member of the Technical Program Committee of the European Solid-State Circuits Conference (ESSCIRC). He is a Senior Member of IEEE. He has successfully been the main supervisor of 14 PhD students to receive their degrees. His research interests include design of radio frequency, microwave, and mm wave integrated circuits, primarily in CMOS technology.
\end{IEEEbiography}

\begin{IEEEbiography}[{\includegraphics[width=1in,height=1.25in,clip,keepaspectratio]{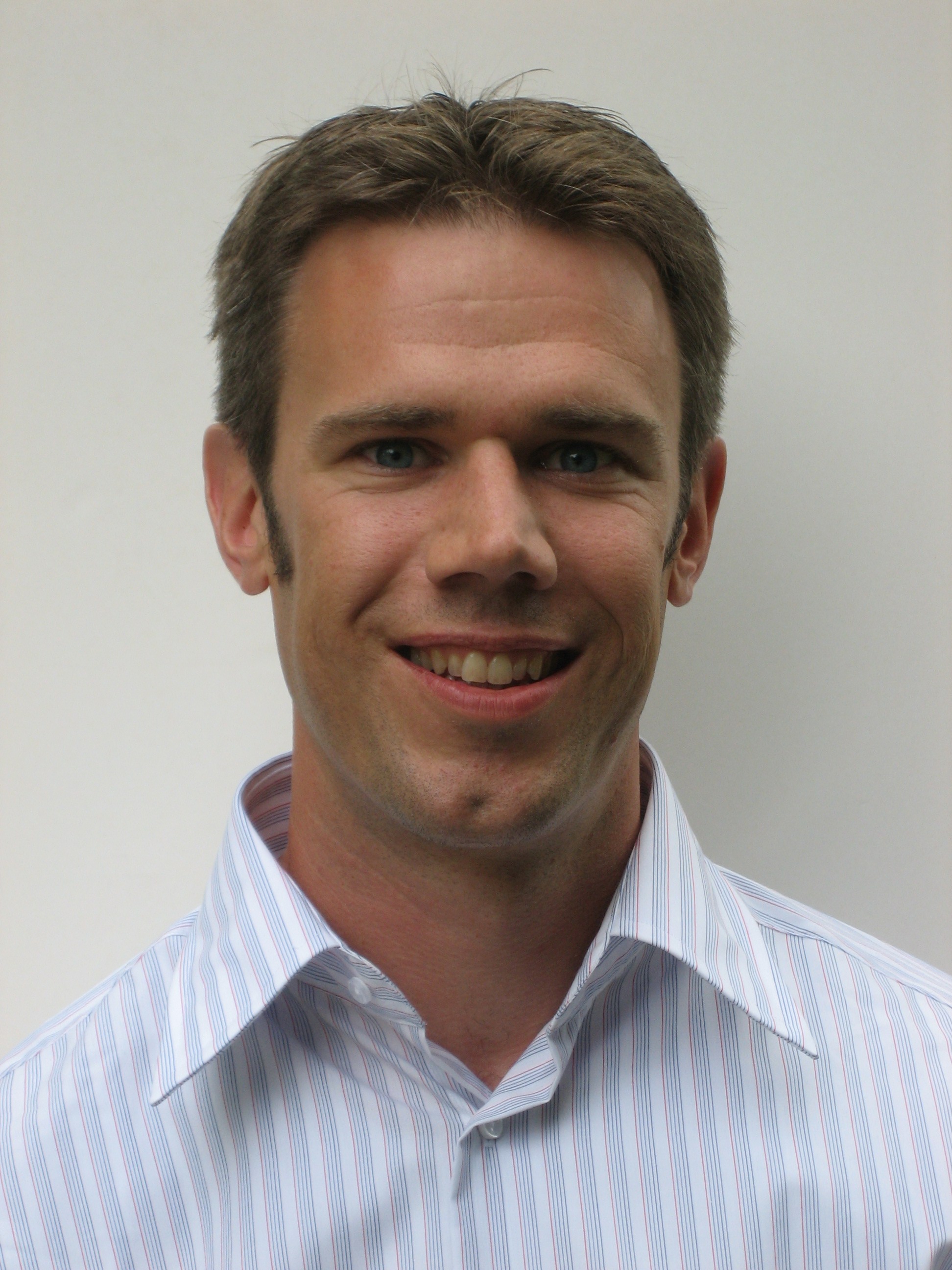}}]{Fredrik Tufvesson} received his Ph.D. in 2000 from Lund University in Sweden. After two years at a startup company, he joined the department of Electrical and Information Technology at Lund University, where he is now professor of radio systems. His main research interest is the interplay between the radio channel and the rest of the communication system with various applications in 5G/B5G systems such as massive MIMO, mmWave communication, vehicular communication and radio-based positioning. Fredrik has authored around 100 journal papers and 150 conference papers, he is fellow of the IEEE and his research has been awarded with the Neal Shepherd Memorial Award for the best propagation paper in IEEE Transactions on Vehicular Technology and the IEEE Communications Society best tutorial paper award.
\end{IEEEbiography}
\balance
\end{document}